%% file: main.tex
\documentclass[sigplan,screen]{acmart}

\setcopyright{acmcopyright}
\acmPrice{15.00}
\acmDOI{10.1145/3519939.3523711}
\acmYear{2022}
\copyrightyear{2022}
\acmSubmissionID{pldi22main-p373-p}
\acmISBN{978-1-4503-9265-5/22/06}
\acmConference[PLDI '22]{Proceedings of the 43rd ACM SIGPLAN International Conference on Programming Language Design and Implementation}{June 13--17, 2022}{San Diego, CA, USA}
\acmBooktitle{Proceedings of the 43rd ACM SIGPLAN International Conference on Programming Language Design and Implementation (PLDI '22), June 13--17, 2022, San Diego, CA, USA}

\input{imports}

\input{macros}

\begin{document}

\title{WebRobot: Web Robotic Process Automation using Interactive Programming-by-Demonstration}

\author{Rui Dong}
\affiliation{
  \institution{University of Michigan, USA}            
  \country{}
}

\author{Zhicheng Huang}
\affiliation{
  \institution{University of Michigan, USA}            
  \country{}
}

\author{Ian Iong Lam}
\affiliation{
  \institution{University of Michigan, USA}            
  \country{}
}

\author{Yan Chen}
\affiliation{
  \institution{University of Toronto, Canada}            
  \country{}
}

\author{Xinyu Wang}
\affiliation{
  \institution{University of Michigan, USA}  
  \country{}
}

\input{abstract}



\maketitle
\input{intro}

\input{overview}

\input{dsl}

\input{problem}

\input{algorithm}

\input{interface}

\input{eval}

\input{related}

\input{ack}

\newpage
\balance
\bibliography{main}

\end{document}

%% file: imports.tex
\usepackage{booktabs}   
\usepackage{subcaption} 

\usepackage{graphicx}
\usepackage{enumitem}
\usepackage[noend]{algpseudocode}
\usepackage{algorithmicx}
\usepackage[boxed,Algorithm]{algorithm}
\usepackage{bbm}
\usepackage{ragged2e}
\usepackage{float}
\usepackage{xcolor}

\usepackage{seqsplit}
\usepackage{hyperref}
\usepackage{url}
\usepackage{xspace}

\usepackage{color}
\usepackage{circledsteps}

\usepackage{adjustbox}
\usepackage{multirow}

\usepackage{array}
\usepackage{makecell}

\usepackage{balance}
\usepackage{lipsum}

\usepackage{rotating}
\usepackage{listings}
\usepackage{wrapfig}    

\usepackage{microtype}

%% file: macros.tex
\newcommand{\toolname}{\textsc{WebRobot}\xspace}
\newcommand{\todo}[1]{{\color{red}#1}\xspace}

\newcommand{\assign}{:=}
\newcommand{\setor}{\cup}

\newcommand{\program}{P}
\newcommand{\block}{S}
\newcommand{\statement}{\block}
\newcommand{\clickinstruction}{\emph{Click}}
\newcommand{\clickstatement}{\clickinstruction}
\newcommand{\scrapetextinstruction}{\emph{ScrapeText}}
\newcommand{\scrapetextstatement}{\scrapetextinstruction}
\newcommand{\scrapelinkinstruction}{\emph{ScrapeLink}}
\newcommand{\scrapelinkstatement}{\scrapelinkinstruction}
\newcommand{\downloadinstruction}{\emph{Download}}
\newcommand{\downloadstatement}{\downloadinstruction}
\newcommand{\sendkeysinstruction}{\emph{SendKeys}}

\newcommand{\senddatainstruction}{\emph{EnterData}}
\newcommand{\senddatastatement}{\senddatainstruction}
\newcommand{\gobackinstruction}{\emph{GoBack}}

\newcommand{\extracturlinstruction}{\emph{ExtractURL}}
\newcommand{\extracturlstatement}{\extracturlinstruction}

\newcommand{\clickaction}{\clickstatement}
\newcommand{\scrapetextaction}{\scrapetextstatement}
\newcommand{\scrapelinkaction}{\scrapelinkstatement}
\newcommand{\downloadaction}{\downloadstatement}

\newcommand{\senddataaction}{\senddatastatement}

\newcommand{\extracturlaction}{\extracturlstatement}

\newcommand{\domchildren}{\emph{Children}}
\newcommand{\domdescendants}{\emph{Dscts}}
\newcommand{\inputchildren}{\emph{ValuePaths}}

\newcommand{\predicate}{\phi}

\newcommand{\domnode}{n}
\newcommand{\domnodes}{N}
\newcommand{\inputvalue}{v}
\newcommand{\inputvalues}{V}
\newcommand{\conststring}{s}
\newcommand{\xpath}{\rho}

\newcommand{\domattribute}{\tau}

\newcommand{\grammareq}{::=}
\newcommand{\foreachloop}[3]{{\color{magenta}{\textbf{\texttt{foreach}}}} \ #1 \ {\color{magenta}{\textbf{\texttt{in}}}} \ #2 \ {\color{magenta}{\textbf{\texttt{do}}}} \ #3}

\newcommand{\while}[1]{{\color{magenta}{\textbf{\texttt{while}}}} \ \emph{true} \ {\color{magenta}{\textbf{\texttt{do}}}} \ #1}
\newcommand{\myif}[2]{{\color{black}{\textbf{\texttt{if}}}} \ #1 \ {\color{black}{\textbf{\texttt{do}}}} \ #2 }

\newcommand{\domnodevar}{\varrho}
\newcommand{\inputpathvar}{\inputvaluevar}
\newcommand{\inputvaluevar}{\vartheta}
\newcommand{\inputvar}{x}

\newcommand{\mydots}{\cdot\cdot}

\newcommand{\resetmycounter}[1]{\setcounter{#1}{0}}

\newcommand{\irule}[2]{\mkern-2mu\displaystyle\frac{#1}{\vphantom{,}#2}\mkern-2mu}

\newenvironment{centermath}{\begin{center}$\displaystyle}{$\end{center}}

\newcommand{\statetrace}{\Pi}
\newcommand{\actiontrace}{A}
\newcommand{\env}{\Sigma}

\newcommand{\evalprogramtop}[3]{#1 \vdash #2 : #3}
\newcommand{\evalprogram}[3]{#1 \vdash #2 \rightsquigarrow #3}
\newcommand{\evaldomnode}[3]{#1 \vdash #2 \rightsquigarrow #3}
\newcommand{\evaldomnodes}[3]{#1 \vdash #2 \rightsquigarrow #3}
\newcommand{\evalinputvalue}[3]{#1 \vdash #2 \rightsquigarrow #3}
\newcommand{\evalinputvalues}[3]{#1 \vdash #2 \rightsquigarrow #3}

\newcommand{\traceconcat}{\statetraceconcat}

\newcommand{\statetraceconcat}{\mkern-0mu \texttt{+} \mkern-1mu \texttt{+} \mkern1mu}
\newcommand{\actiontraceconcat}{\statetraceconcat}
\newcommand{\emptylist}{\texttt{[]}}
\newcommand{\mystate}{\pi}
\newcommand{\action}{a}

\newcommand{\valid}{\emph{valid}}

\newcommand{\inputvaluepath}{\theta}

\newcommand{\listconcat}{\mkern-2mu :: \mkern-2mu}

\newcommand{\inputdata}{I}

\newcommand{\domtag}{t}

\newcommand{\xpathpredicate}{\predicate}

\newcommand{\myprocedureindent}[3]{
\Statex{\hspace{#3} {\bf procedure} \textsc{#1} (#2)}
}

\algnewcommand\Input{\hspace{-12pt}\textbf{input: }}
\algnewcommand\Output{\hspace{-12pt}\textbf{output: }}
\algnewcommand\Invariant{\hspace{-12pt}\textbf{invariant: }}
\algrenewcommand\algorithmicindent{1em}

\newcommand{\vecstatetrace}{\vec{\statetrace}}
\newcommand{\vecactiontrace}{\vec{\actiontrace}}

\newcommand{\sublisteq}{\sqsubseteq}

\newcommand{\statements}{\widetilde{\statement}}

\newcommand{\rewrites}{W}
\newcommand{\srewrites}{\Omega}

\newcommand{\worklist}{W}
\newcommand{\setprograms}{\widetilde{P}}

\newcommand{\unifyop}{\circledast}
\newcommand{\unify}[3]{\vdash #1 \unifyop #2 \twoheadrightarrow #3}
\newcommand{\unifyvar}[4]{#1 \vdash #2 \unifyop #3 \twoheadrightarrow #4}

\newcommand{\parameterize}[3]{#1 \vdash #2 \twoheadrightarrow #3}

\newcommand{\sslash}{\mkern-0mu/\mkern-5mu/\mkern-0mu}

\newcommand{\doubleprime}{'\mkern-2mu'}
\newcommand{\tripleprime}{'\mkern-2mu'\mkern-2mu'}

\newcommand{\helena}{Helena\xspace}

\newcommand{\invariant}{\mathcal{I}}

\newcommand{\etc}{\emph{etc}}
\newcommand{\ie}{{i.e.}}

\newcolumntype{L}[1]{>{\raggedright\let\newline\\\arraybackslash\hspace{0pt}}m{#1}}
\newcolumntype{C}[1]{>{\centering\let\newline\\\arraybackslash\hspace{0pt}}m{#1}}
\newcolumntype{R}[1]{>{\raggedleft\let\newline\\\arraybackslash\hspace{0pt}}m{#1}}

\newenvironment{myindent}
{\par\leftskip0.4cm\relax\rightskip0.4cm\relax}
{\par\leftskip0cm\relax\rightskip0cm\relax}

\newcommand{\egg}{\texttt{egg}\xspace}

%% file: abstract.tex
\begin{abstract}
\vspace{-1pt}

It is imperative to democratize robotic process automation (RPA), as RPA has become a main driver of the digital transformation but is still technically very demanding to construct, especially for non-experts. 
In this paper, we study how to automate an important class of RPA tasks, dubbed \emph{web RPA}, which are concerned with constructing software bots that automate interactions across data and a web browser. 
Our main contributions are twofold. 
First, we develop a formal foundation which allows semantically reasoning about web RPA programs and formulate its synthesis problem in a principled manner. 
Second, we propose a web RPA program synthesis algorithm based on a new idea called \emph{speculative rewriting}. 
This leads to a novel \emph{speculate-and-validate} methodology in the context of rewrite-based program synthesis, which has also shown to be both theoretically simple and practically efficient for synthesizing programs from demonstrations. We have built these ideas in a new interactive synthesizer called \toolname and evaluate it on 76 web RPA benchmarks. 
Our results show that \toolname automated a majority of them effectively. 
Furthermore, we show that \toolname compares favorably with a conventional rewrite-based synthesis baseline implemented using \egg. 
Finally, we conduct a small user study demonstrating \toolname is also usable.

\end{abstract}

%% file: intro.tex
\section{Introduction}
\label{sec:intro}

\begin{figure}
\centering
\includegraphics[width=0.47\textwidth]{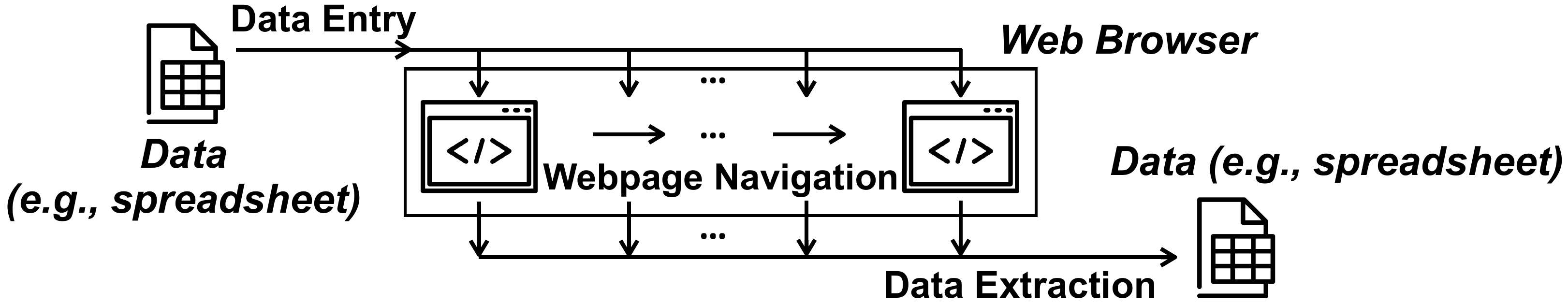}
\vspace{-5pt}
\caption{Illustration of web RPA.}
\vspace{-5pt}
\label{fig:webrpa}
\end{figure}

\begin{figure}
\centering
\includegraphics[width=0.42\textwidth]{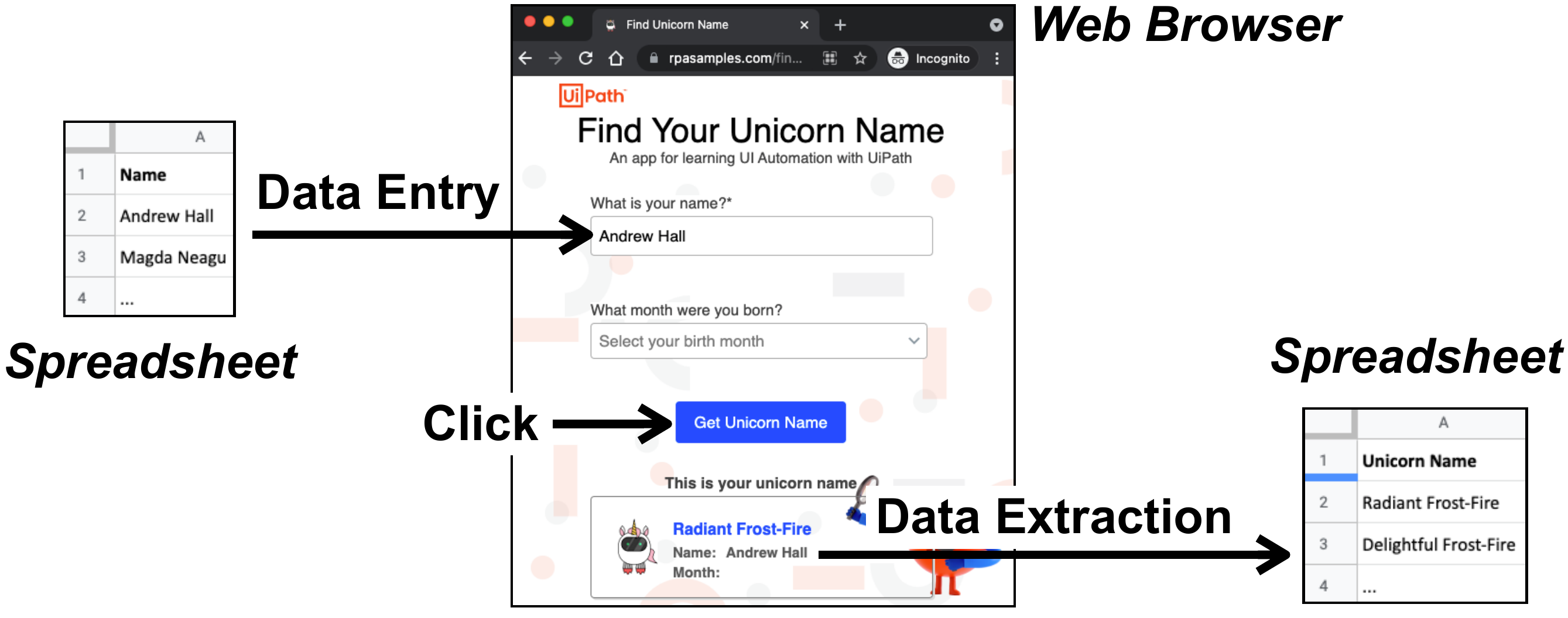}
\vspace{-5pt}
\caption{A real-world web RPA problem from UiPath.}
\label{fig:ex1}
\vspace{-5pt}
\end{figure}

Robotic process automation (RPA) is a software technology that aims to streamline the process of creating \emph{software robots}
that emulate user interactions with digital applications such as web browsers and spreadsheets~\cite{rpa-1,rpa-2,fischer2021diy,leno2021discovering,leno2020robidium,zhang2020process,wewerka2020robotic,agostinelli2020towards}. 
These robots are essentially programs: like humans, they can perform tasks such as entering data, completing keystrokes, navigating across pages, extracting data, \etc. However, they, once programmed, can perform tasks much faster with fewer mistakes. Therefore, RPA has the potential to significantly simplify business workflows and improve the productivity for \emph{both organizations and individuals}~\cite{uipath-studiox-webinar-slides}. Gartner predicted that RPA will remain the fastest-growing software market in the next several years~\cite{gartner-report-2021}.

While RPA has become a main driver of the digital transformation, it is still technically very demanding to construct automation programs, and consequently, not everyone can build software robots that suit their needs. For instance, it is estimated that 3-7\% of tasks deemed important \emph{by an organization} have been automated, whereas a long tail of more than 40\% of \emph{individual-driven} tasks yet are still to be automated~\cite{uipath-studiox-webinar-slides}. 
These tasks represent a high percentage of automation opportunities to scale RPA to individual non-expert end-users.

\textbf{\em Web RPA.}
How to democratize RPA in order to foster its adoption 
among non-experts is a broad, new but increasingly important problem. In this paper, we consider an important subset of RPA tasks, dubbed \emph{web RPA}, and investigate how to automate this class of tasks. 
As illustrated in Figure~\ref{fig:webrpa}, web RPA involves \emph{interactions between data and a web browser}. 
For example, it involves \emph{programmatically} entering  data (in a semi-structured format), extracting data from webpages, as well as navigating across multiple webpages. 
Conceptually, web RPA is close to web/browser automation, where the key distinction is that RPA emphasizes \emph{interactions across/within applications}, while browser automation has to do with \emph{web browsers}.
In other words, one can view web RPA as the ``intersection'' of RPA and web automation; hence the name.\footnote{Web automation is a broad term. We note that web RPA is highly related to web automation but in this work, we do not precisely distinguish them.}

Let us consider a real-world scenario (shown in Figure~\ref{fig:ex1}) from a recent webinar~\cite{uipath-studiox-webinar-video} by UiPath, a leading RPA company. 
In this example, a manager working for a unicorn adoption agency wanted to test a hypothesis that sending follow-up emails with their unicorn names to customers would increase the adoption rate. Unfortunately, the customer relationship management system is disconnected from the web-based unicorn name generator. In other words, there is no easy way to automatically generate a unicorn name for each customer. The manager tried to seek help from the IT but was told that creating an automation program for this job is expensive unless there is a provably sufficient return on it. In the end, they had to manually perform this experiment: export customer information into a spreadsheet, \emph{copy-paste every name from the sheet, enter it in the unicorn generator, scrape the generated name for each customer,} and finally send emails with unicorn names. This is very tedious. A key problem in this process is how to create a program that interacts with the web-based generator and the spreadsheet in order to create names for all customers.  
This is exactly a web RPA problem.\footnote{This example involves one single webpage but we have many benchmarks that involve navigating across multiple pages (see Section~\ref{sec:eval}).}

Web RPA sits at the intersection of multiple areas, such as programming languages and human-computer interaction. While it has been studied in different forms by different communities, to the best of our knowledge, there is no principled approach that automatically generates web RPA programs in a comprehensive manner. 
For instance, while being able to scrape data across webpages, \helena~\cite{chasins2019democratizing} has relatively less support for programmatic data entry. Furthermore, it may generate wrong programs which, in our experience, are not always easy to ``correct'' using \helena's build-in features.
On the other hand, the HCI and databases communities have proposed various interfaces~\cite{leshed2008coscripter,lin2009end} and wrapper induction techniques~\cite{raza2020web,anton2005xpath,gulhane2011web}, which are even more restricted and can automate only single-webpage tasks. 
Finally, while some ``low-code'' solutions based on record-and-replay exist on the industrial market  (such as iMacros~\cite{imacros}), they require significant manual efforts (e.g., adding loops), which makes them potentially less accessible to non-expert end-users.

\begin{figure}
\centering
\includegraphics[width=0.4\textwidth]{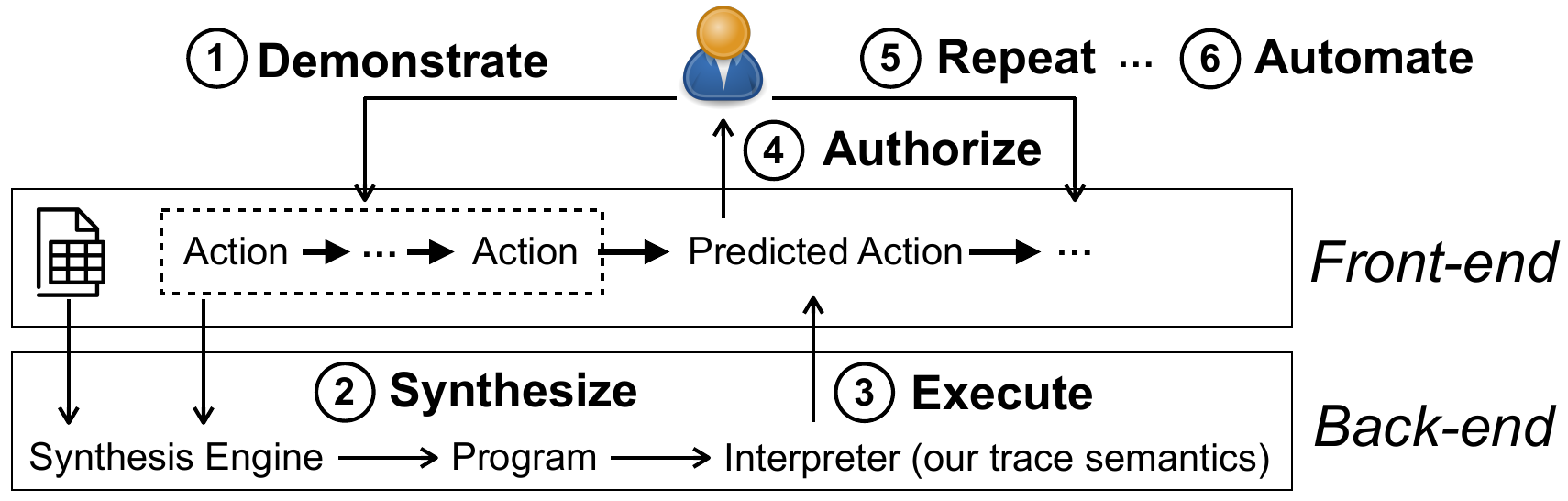}
\vspace{-8pt}
\caption{Schematic workflow of our approach.}
\label{fig:workflow}
\vspace{-12pt}
\end{figure}

\textbf{\em Interactive programming-by-demonstration (PBD) for web RPA.}
Our first contribution is a new approach that automates web RPA tasks from demonstrations \emph{interactively}.  
Compared to existing work, our approach is more automated, resilient to ambiguity, and applicable for web RPA. 
Figure~\ref{fig:workflow} shows the schematic workflow of our approach. 
To automate a task, the user just needs to perform it as usual but using our interface (step {\small \Circled{1}}). All the user-demonstrated \emph{actions} are recorded and sent to our back-end synthesis engine. Then, we synthesize a program $\program$ that ``satisfies'' the demonstration (step {\small \Circled{2}}). That is, $\program$ is guaranteed to \emph{reproduce} the recorded actions, but $\program$ may also \emph{produce more actions afterwards}. We then ``execute'' $\program$ to produce an action that the user \emph{may want to perform next} and visualize this predicted action via our interface (step {\small \Circled{3}}). Finally, the user inspects the prediction and chooses to accept or reject it (step {\small \Circled{4}}). This \emph{interactive} process repeats until there is sufficient confidence that the synthesized program is intended (step {\small \Circled{5}}); after that, it will take over and automate the rest of the task (step {\small \Circled{6}}). Note that, if at any point the user spots anything abnormal, they can  still interrupt and enter the demonstration phase again.

We highlight several salient features of our approach. 
First, it is \emph{automated}: users only need to provide demonstrations, without needing to write programs. 
Second, it is \emph{interactive}: whenever a synthesized program is not desired, the user can simply interrupt and continue demonstrating more actions, without having to edit programs. 
Finally, it could synthesize programs effectively from an expressive language, thanks to \emph{a systematic problem formulation} and \emph{a new search algorithm.}

\textbf{\em Systematic formulation of PBD for web RPA.} 
Our second contribution is a systematic formulation for the problem of synthesis from \emph{action-based demonstrations.} 
In particular, the question we aim to address here is: what does it mean for a program to \emph{satisfy a trace of user-demonstrated actions}?
This problem is extremely understudied in a formal context.
To the best of our knowledge, the latest work to date is the seminal work~\cite{lau1998programming,lau2001programming,lau2003programming} by Tessa Lau and their co-authors in the 1990s. 
However, Lau's work considers \emph{state-based} demonstrations; that is, in their work, a demonstration is defined as a sequence of program states. In contrast, our work concerns \emph{action-based} demonstrations---a demonstration is a trace of actions. 
In this context, we are not aware of any prior work that has formalized the \emph{semantic} notion of satisfaction. 
As a result, existing techniques~\cite{mo1990learning,masui1994repeat,chasins2018rousillon,chasins2019democratizing} resort to heuristics and task-specific rules to detect patterns in the action trace in order to generalize it to programs with loops.
In this work, we formulate the action-based PBD problem by formalizing the \emph{trace semantics} for an expressive web RPA language. 
In a nutshell, our semantics ``executes'' a program (with loops) and produces its ``execution trace'' of \emph{actions} by unrolling loops and replacing variables with values. 
Therefore, with our semantics, we can now \emph{check} a program against a trace of actions. 
Furthermore, this semantics also plays a pivotal role in our \emph{search} algorithm, which we will explain next.

\textbf{\em Action-based PBD using speculative rewriting.} 
Once we can check a program against a trace of actions, the next question we ask is: how to \emph{search} for programs that satisfy the given trace? 
This brings us to the third contribution of our work, which is a novel rewrite-based synthesis algorithm based on a new idea called \emph{speculative rewriting}. 
The basic idea is simple: we rewrite a slice of the trace into a (one-level) loop which produces that slice, using a set of predefined rules; if we do this iteratively, we can generate nested loops from the inside out. 
The issue is, it is very hard to define a \emph{complete} set of \emph{correct-by-construction} rules in our domain, because our trace may result from executing loops (from an arbitrary program) for arbitrarily many times. 
In other words, pattern-matching the \emph{entire} trace in a purely rule-based manner does not scale. 
In order to scale to complex programs, our idea is to combine \emph{rule-based pattern-matching} and \emph{semantic validation} in the rewrite process via an intermediate \emph{speculation} step.
More specifically, instead of pattern-matching \emph{all iterations} to directly generate \emph{true rewrites}, our idea is to pattern-match \emph{a couple of iterations} and generate \emph{speculative rewrites}, or \emph{s-rewrites}.
While an s-rewrite might not be a true rewrite in general, they \emph{over-approximate} the set of true rewrites and are much easier to generate.
We then use our \emph{trace semantics} to \emph{validate} s-rewrites and retain only those true rewrites.

Our method is closely related to two lines of work. First, it builds upon the ``guess-and-check'' idea introduced by the counterexample-guided inductive synthesis (CEGIS) framework~\cite{solar2008program}, but we show how to extend this idea for rewrite-based synthesis beyond the traditional application scenarios with example-based and logical specifications. Second, our method incorporates the idea of \emph{semantic rewrite rules} from recent work~\cite{nandi2020synthesizing,willsey2021egg}, but we augment this standard correct-by-construction, rule-based rewrite approach with a novel guess-and-check step.\footnote{We will elaborate on this in the remainder of this paper.} 
We found this new idea to be both theoretically simple and practically efficient in our domain.
We also believe this methodology is potentially useful in the more general context of rewrite-based synthesis and in other problem domains with similar trace generalization problems. 

\textbf{\em Human-in-the-loop interaction model.}
As a proof-of-concept, we have also developed a user interface to facilitate user interactions with our synthesizer. 
Our interface combines programming-by-demonstration, action visualization, and interactive authorization within a human-in-the-loop model,  which has shown to be useful in practice for reducing the gulfs of execution and evaluation~\cite{norman2013design}.

\textbf{\em Implementation and evaluation.}
We have implemented our proposed ideas in a tool called \toolname and evaluate it across four experiments. 
First, we evaluate \toolname's synthesis engine on 76 real-world web RPA benchmarks and show that it can synthesize programs effectively. 
Second, we perform an ablation study and show that all of our proposed ideas are important. 
Furthermore, we conduct a user study with eight participants which shows that \toolname can be used by non-experts. 
Finally, we compare \toolname with a rewrite-based synthesis approach and our results show that \toolname significantly advances the state-of-the-art.

\vspace{1pt}
In summary, this paper makes the following contributions: 
\begin{itemize}[leftmargin=*]
\vspace{-2pt}
\item 
We identify the web RPA program synthesis problem. 
\item 
We formalize a trace semantics of our web RPA language, laying the formal foundation for its synthesis problem. 
\item 
We present a novel programming-by-demonstration algorithm based on a new idea called speculative rewriting. 
\item 
We develop a new human-in-the-loop interaction model. 
\item 
We implement our ideas in a new tool called \toolname. 
\item 
We evaluate \toolname on 76 tasks and via a user study. 
\end{itemize}

%% file: overview.tex
\section{Overview of \toolname}
\label{sec:overview}

In this section, we highlight some key features of \toolname using a motivating example\footnote{\url{https://forum.imacros.net/viewtopic.php?f=7&t=21028}} from the iMacros forum. 

\begin{figure}
\centering
\includegraphics[width=0.36\textwidth]{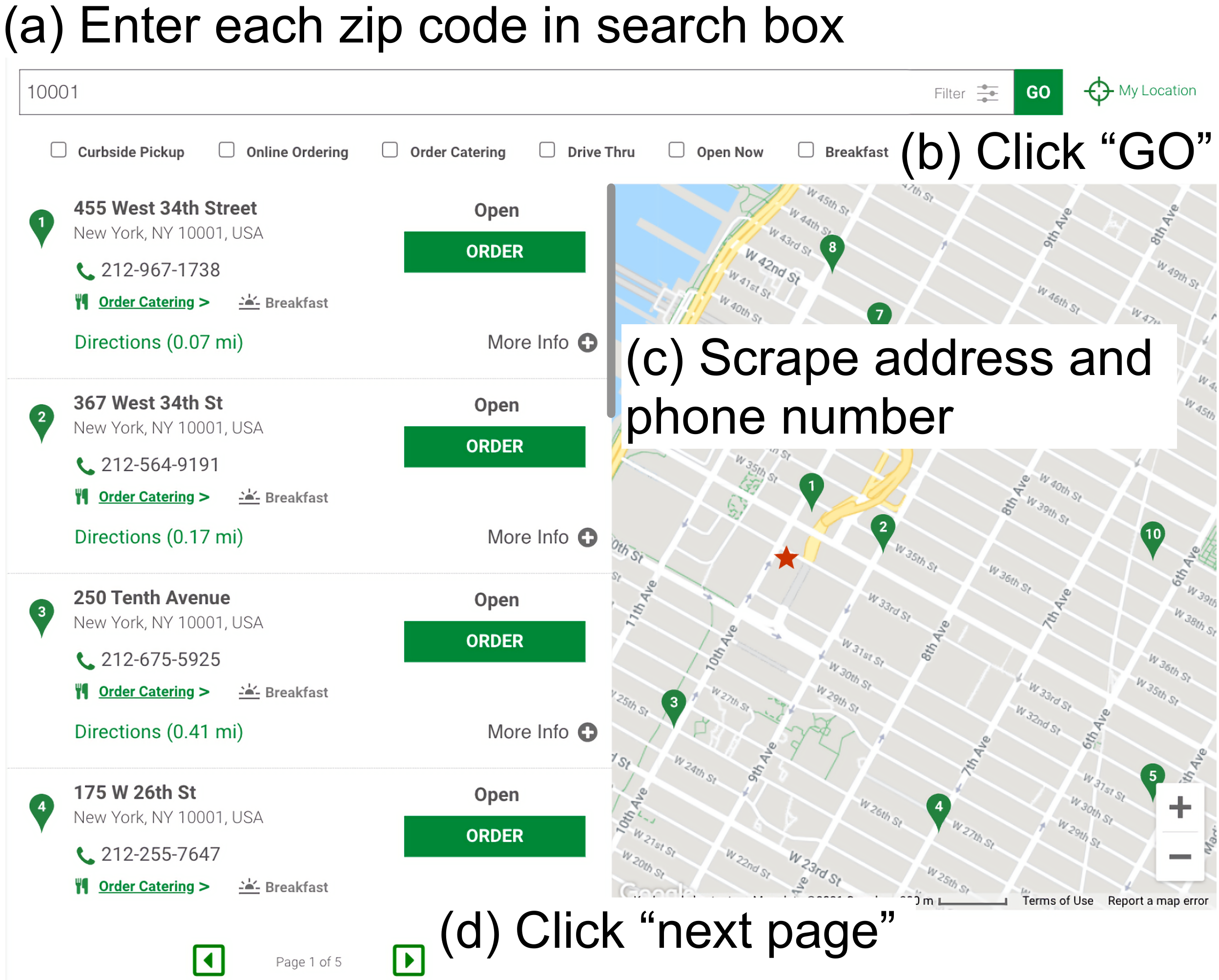}
\vspace{-5pt}
\caption{A motivating example: scrape address and phone number for all stores across all pages and for all zip codes.}
\label{fig:motivating}
\vspace{-15pt}
\end{figure}

\textbf{\em Motivating example.}
Given a list of zip codes, Ellie wants to extract store information from the Subway website\footnote{\url{http://www.subway.com/storelocator/}}. Since Ellie is not familiar with programming, she has to \emph{manually} perform this task (shown in Figure~\ref{fig:motivating}): (a) enter the first zip in the search box, (b) click the search button which then shows five pages of search results, (c) scrape store information on the first page, (d) click the ``next page'' button and repeat this process for all pages and for all zip codes. 

\begin{figure*}
\centering
\includegraphics[scale=0.37]{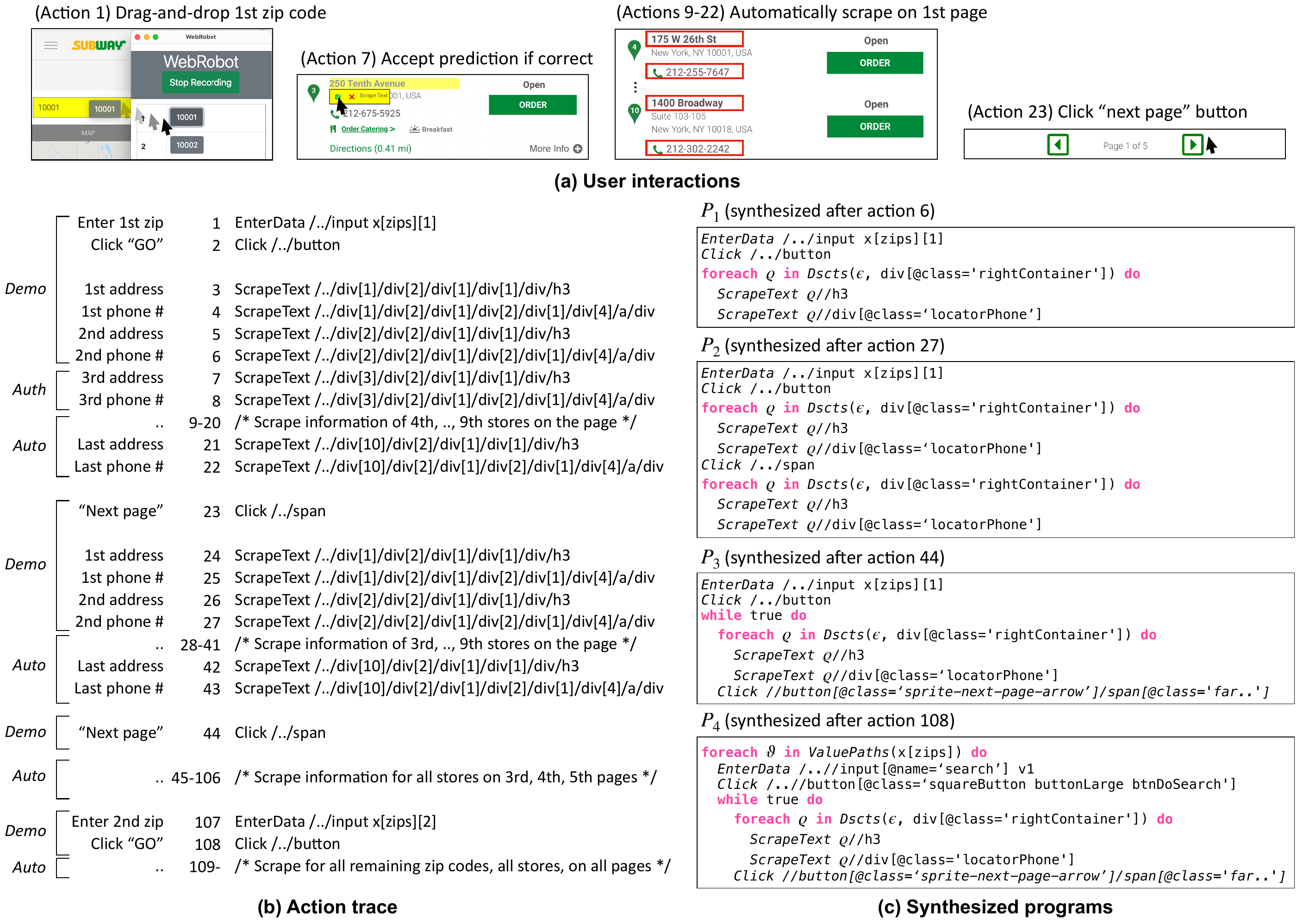}
\vspace{-10pt}
\caption{
(a) User interactions with the web browser and \toolname. 
(b) Action trace recorded by \toolname, where a short explanation is attached to the left of each action. 
(c) Programs synthesized by \toolname at different points.}
\label{fig:demo}
\vspace{-5pt}
\end{figure*}

\textbf{\em \toolname.}
Ellie could use our tool to automate this task. Once \toolname is fired up, Ellie would first import the list of zip codes and then  perform the task using \toolname. This process is illustrated in Figure~\ref{fig:demo}(a). 
In particular, Ellie first drags the first zip and drops it in the search bar (action 1). Then, she clicks the GO button and starts scraping information of the first two stores on the first page: see actions 2-6 in Figure~\ref{fig:demo}(b), although we do not show them in Figure~\ref{fig:demo}(a). All these actions are recorded by \toolname in an action trace, which is shown in Figure~\ref{fig:demo}(b). After six actions, \toolname is able to synthesize a program $\program_1$, as shown in  Figure~\ref{fig:demo}(c), which extracts the address and phone number for each store on the first page. Next, \toolname performs an interactive ``authorization'' step: it executes $\program_1$ to produce the next action which is then visualized to Ellie (see Figure~\ref{fig:demo}(a), action 7). This is correct, so Ellie accepts it. After a couple of rounds, \toolname takes over and automates the scraping work on the first page (Figure~\ref{fig:demo}(a), actions 9-22).

\toolname would terminate after action 22. Thus, Ellie needs to click the ``next page'' button and extract information for a couple of stores on the second page. These actions are also recorded; see Figure~\ref{fig:demo}(b), actions 23-27. At this point, \toolname infers a different program $\program_2$ which has two loops one after another, where the second loop extracts information of all stores on the second page. Using $\program_2$, \toolname is able to automatically scrape the second page; however, it terminates, again, right before the ``next page'' button. 

This time, once Ellie clicks ``next page'' (\ie, action 44), we can synthesize $\program_3$--see Figure~\ref{fig:demo}(c)--which contains an outer \emph{while} loop that first uses an inner loop for scraping and then clicks ``next page'' at the end. $\program_3$ now is able to automatically scrape all store information for the remaining pages. 

Since Ellie needs to repeat this scraping process for all zip codes, she will enter the second zip code and click ``GO'' again (actions 107-108), after which \toolname can synthesize $\program_4$ that has a three-level loop. $\program_4$ first iterates over all zip codes in the given list and then uses a doubly-nested loop to scrape across all pages. At this point, Ellie is done. 

\vspace{1pt}
We highlight some salient features of \toolname below.

\textbf{\em Interactive PBD to resolve ambiguity.} 
\toolname does not need users to provide multiple small demonstrations as in traditional PDB approaches~\cite{lau2001programming}; instead, it synthesizes programs \emph{while} the user is performing the task. 
In case the synthesized program is not intended, \toolname does not ask the user for an edit on the program as in program-centric tools (such as \helena). Rather, it allows the user to take over and correct the behavior. For instance, in the ``authorization'' phase, \toolname visualizes potentially multiple options for the next action and let the user select the one that is desired. 
This design aims to facilitate interactive disambiguation. 

\textbf{\em Satisfaction check using trace semantics.}
Consider $\program_1$ from Figure~\ref{fig:demo}(c) which is synthesized from the first 6 actions $\action_1, \mydots, \action_6$ in Figure~\ref{fig:demo}(b). In other words, $\program_1$ \emph{satisfies} the action trace $[ \action_1, \mydots, \action_6 ]$. 
To perform this satisfaction check, we \emph{simulate} the execution of $\program_1$ using our trace semantics. Note that this is a simulated execution, rather than actually executing $\program_1$ in the browser, because during the synthesis process, programs might have side-effects that are not intended. 
Our trace semantics would first execute the $\senddatainstruction$ and $\clickinstruction$ statements before the loop, which essentially reproduces $\action_1, \action_2$. Then, we unroll the loop twice, reproducing $\action_3, \mydots, \action_6$. A subtle aspect here is that the actions produced by our simulated execution might not be \emph{syntactically} the same as those in the recorded trace, since $\program_1$ might use selectors that are different from those in the recorded action trace. 
Thus, we check if an action produced by our semantics and an action in the demonstrated trace refer to the same Document Object Model (DOM) node; this is done by also recording a trace of DOMs in tandem. We will explain this in detail in Section~\ref{sec:dsl}.

\textbf{\em Rewrite-based PBD.}
How to synthesize programs from an action trace? 
We take a rewrite-based approach. 
Consider the trace $[ \action_1, \mydots, \action_6 ]$ and $\program_1$ in our previous example: the loop in $\program_1$ is rewritten from actions $ \action_3, \mydots, \action_6$. 
\toolname can also synthesize nested loops. 
For instance, the inner loop in $\program_3$ corresponds to multiple \emph{slices} of actions, such as 3-22, 24-43. Once identified, these slices are rewritten to (multiple occurrences of) the same loop. Then, \toolname will generate a nested loop from the inside out by essentially treating the (inner) loop as one action and rewriting again. In this case, it rewrites actions 3-106 to the \emph{while} loop in $\program_3$.

\textbf{\em Selector search.}
In addition to identifying iteration boundaries, \toolname also considers other selectors, beyond full XPath expressions that are recorded in the trace, since the desired program may not use those recorded. For example, $\action_4$ in Figure~\ref{fig:demo}(b) is a full XPath, whereas the corresponding statement in $\program_1$ (namely, the second statement in the loop) uses a more general selector (with \emph{div[@class='locatorPhone']}). 
Considering alternative selectors allows to induce more general programs, but it also makes the problem more challenging.

\textbf{\em Speculative rewriting.}
A standard rewrite-based synthesis approach requires a set of \emph{correct-by-construction} rewrite rules~\cite{nandi2020synthesizing, willsey2021egg}, meaning they always generate sound rewrites. 
In our domain, if we follow this idea, we need to design rules that pattern-match actions which result from an unknown number of loop iterations and from arbitrarily complex loop structures; this is hard to scale to complex web RPA tasks. 
\emph{Our idea is to pattern-match actions from a couple of iterations.} 
For instance, given $[\action_1, \mydots, \action_{22}]$ in Figure~\ref{fig:demo}(b) that corresponds to $\program_1$, instead of pattern-matching $\action_3, \mydots, \action_{22}$, using rules, to synthesize a \emph{true} rewrite (\ie, a loop), we pattern-match $\action_3, \mydots, \action_6$ and \emph{speculate} a potential rewrite $\program_l$, assuming $\action_3, \mydots, \action_6$ ``come from'' the first two iterations of $\program_{\emph{l}}$. 
This is conceptually simpler and faster, but the downside is that $\program_{\emph{l}}$ might not be a true rewrite, since it is inferred from only the first two iterations.

\textbf{\em Semantic validation.}
Our next idea is to check if a speculative rewrite  (or, s-rewrite) $\program_l$ is a true rewrite by executing $\program_l$ under our trace semantics. 
The goals are twofold. First, we check if $\program_l$ can rewrite a \emph{longer} slice of actions beyond the first iteration; if not, we filter out $\program_l$. 
Second, if $\program_l$ indeed rewrites beyond the first iteration, semantic validation also gives us the (longer) slice that $\program_l$ could rewrite.
As we can see, building a formal semantic foundation allows us to not only systematically formulate the synthesis problem for web RPA, but also develop an effective algorithm to solving it.

%% file: dsl.tex
\vspace{-5pt}
\section{Web RPA Language and Trace Semantics}
\label{sec:dsl}

This section lays the formal foundation for web RPA.

\vspace{-5pt}
\subsection{Syntax}
\vspace{-2pt}

Our syntax is shown in Figure~\ref{fig:dsl}.
Intuitively, a program $\program$ in this language is a sequence of statements that emulates user interactions with a web browser and a data source. Its input variable $\inputvar$ is a data source $\inputdata$ represented in JSON-like format:  
\vspace{-8pt}
\[\small 
\arraycolsep=1pt\def\arraystretch{1}
\begin{array}{lll}
\inputdata \ \ & \grammareq & \{ \emph{key} : \emph{value}, \ \mydots, \ \emph{key} : \emph{value} \} \\ 
\emph{key} \ \ & \grammareq & \emph{string} \ \ \ \ \ \ \ \ 
\emph{value} \grammareq \emph{string} \ | \ \emph{integer} \ | \ \inputdata \ | \ [ \emph{value}, \mydots, \emph{value} ]  \\ 
\end{array}
\vspace{-1pt}
\]
This allows using any semi-structured data as our data source.

A statement $\statement$, in the simplest case, performs an action on the current webpage. For example, $\clickstatement$ clicks a DOM node located by a selector $\domnode$. $\scrapetextinstruction$ scrapes the text inside a node specified by $\domnode$. 
Some statements are parameterless (e.g., $\gobackinstruction$ that goes back to the previous page and $\extracturlinstruction$ that gives the URL of current webpage). 
Some statements might take multiple parameters. For example, $\sendkeysinstruction$ types a constant string $\conststring$ into an editable field given by a selector $\domnode$. $\senddatainstruction$ enters a value $\inputvalue$ from input data $\inputdata$ to a field located by $\domnode$. Note that $\inputvalue$ is represented using a \emph{value path}, which is essentially a sequence of keys and array indices in order to access the value from $\inputdata$. 
On the other hand, a selector $\domnode$ in our language is essentially an XPath expression~\cite{xpath-1} but it may contain a variable $\domnodevar$ \emph{at the beginning}. In particular, $\domnode\slash\predicate[i]$ gives the $i$-th \emph{child} of a DOM node $\domnode$ that satisfies a predicate $\predicate$. $\domnode\sslash\predicate[i]$ gives the $i$-th \emph{descendant} which satisfies $\predicate$ among all nodes in the subtree rooted at $\domnode$. Our language has multiple types of predicates. The simplest one is an HTML tag $\domtag$. For instance, $\domnode/\emph{span}[1]$ returns the first child of $\domnode$ with tag $\emph{span}$. The next predicate $\domtag[@\domattribute=\conststring]$ means the desired DOM node should have tag $\domtag$ and its attribute $\domattribute$ should take value $\conststring$. For instance, $\domnode\sslash\emph{div}[@\emph{class}=``a"][2]$ returns the second descendant of $\domnode$ that has tag $\emph{div}$ and whose \emph{class} attribute value is string ``\emph{a}''.

\begin{figure}

\footnotesize
\centering
\[
\begin{array}{llll}

\emph{Program} & 
\program & 
\grammareq & 
\block; \mydots; \block \\

\emph{Statement} & 
\block & 
\grammareq & 
\clickinstruction( \domnode) \ | \ 
\scrapetextinstruction( \domnode ) \ | \  
\scrapelinkinstruction( \domnode ) \\ 
& & \ \ \ | & 
\downloadinstruction( \domnode) \ | \ 
\gobackinstruction \ | \ 
\extracturlinstruction \\ 
& & \ \ \ | & 
\sendkeysinstruction( \domnode, \conststring ) \ | \ 
\senddatainstruction( \domnode, \inputvalue ) \\ 
& & \ \ \ | & 
\foreachloop{\domnodevar}{\domnodes}{ \program } 
\ \ \ \  \ \ \ \ \ \ \ \ \ \ \ \ \ \emph{(selectors loop)}
\\ 
& & \ \ \ | & 
\foreachloop{\inputvaluevar}{\inputvalues}{\program} 
\ \ \ \  \ \ \ \ \ \ \ \ \ \ \emph{(value path loop)}
\\ 
& & \ \ \ | & 
\while{\{ \program; \clickstatement(\domnode)
\} }
\ \ \ \ \ \ \ \ \ \ \ \ \emph{(while loop)}
\\

\emph{Selector} & 
\domnode & 
\grammareq & 
\epsilon \ | \ 
\domnodevar \ | \ 
\domnode/\predicate[i] \ | \ 
\domnode\sslash\predicate[i] \\

\emph{Value Path} & 
\inputvalue & 
\grammareq & 
\inputvar \ | \ 
\inputvaluevar \ | \ 
\inputvalue[\emph{key}] \ | \ 
\inputvalue[i] \\

\emph{Selectors} & 
\domnodes & 
\grammareq & 
\domchildren( \domnode, \predicate) \ | \ 
\domdescendants( \domnode, \predicate) \\

\emph{Value Paths} & 
\inputvalues & 
\grammareq & 
\inputchildren( \inputvalue ) \\

\emph{Predicate} & 
\predicate & 
\grammareq & 
\domtag \ | \ 
\domtag[@\domattribute=\conststring] \\

\end{array}
\]

\vspace{-10pt}
\[
\hspace{4pt}
\begin{array}{llll}
\conststring \grammareq \emph{string} \ \ \ \ \ 
i \grammareq \emph{integer} \ \ \ \ \ 
\domtag \grammareq \emph{HTML tag} \ \ \ \ \ 
\domattribute \grammareq \emph{HTML attribute} 
\end{array}
\]
\vspace{-18pt}
\caption{Syntax of our web RPA language.}
\label{fig:dsl}
\vspace{-10pt}
\end{figure}

Statements could also be loopy. The first type of loop is \emph{selector loops} which iterate over a list $\domnodes$ of DOM nodes on a webpage. This construct is to emulate loopy user interactions on webpages, such as scraping a list of elements. In particular, $\domnodes$ returns a list of selectors. During the $i$-th iteration, the loop variable $\domnodevar$ binds to the $i$-th selector in $\domnodes$ under which the loop body $\program$ gets executed. Note that a statement in $\program$ could use $\domnodevar$ and it may also be a loop. The next loop type is \emph{value path loops}, which are used to emulate loopy interactions with input data. In this case, $\inputvalues$ evaluates to a list of value paths, $\inputvaluevar$ binds to each value path, and $\program$ is executed in this context. Our last type of loops is \emph{while loops}, where the termination condition is that the DOM node $\domnode$ in the last $\clickstatement$ statement no longer exists on the webpage. This construct is primarily used to handle pagination where the user needs to repeatedly click the ``next page'' button until there is no next page.

\subsection{Trace Semantics}

So far, we have seen the DSL syntax, which is new but fairly standard. 
Now, in this section, we will formalize its semantics, which is a key distinction of our paper from prior work. 

\textbf{\em Design rationale.}
Let us first briefly present our thought process in designing this semantics. 
Recall that a program $\program$ in our language takes as input a data source $\inputdata$ and is executed on an initial DOM; $\program$ has side-effects that change the DOM, eventually terminating at some browser state. Thus, one may start with the following semantics definition. 
\vspace{-2pt}
\[
\begin{array}{c}
\evalprogramtop{ \mystate, \env }{\program}{ \mystate' } 
\end{array}
\vspace{-3pt}
\]
Here, $\mystate$ is the initial DOM, $\env$ is an environment that tracks variable values, and $\mystate'$ is the final DOM when $\program$ terminates. 

However, there is a gap between this semantics and our specification: one is based on DOMs and the other is based on actions. 
This brings us to \emph{our first key insight}: the semantics (for our synthesis technique) should incorporate actions that the program executes. This is primarily because in synthesis, we typically use semantics to validate candidate programs against the specification. 
This leads to the following design. 
\[
\begin{array}{c}
\evalprogramtop{ \mystate, \env }{\program}{ \actiontrace', \mystate' } 
\end{array}
\vspace{-2pt}
\]
The key idea in this design is to track the trace $\actiontrace'$ of actions taken by $\program$ during its execution. 
Here, $\actiontrace'$ is a list $[\action_1, \mydots, \action_m]$ of actions where an action is defined as follows. 
\[\small
\arraycolsep=1pt\def\arraystretch{1.1}
\begin{array}{lll}
\action & \grammareq  \clickaction( \xpath ) \ | \  \scrapetextaction( \xpath ) \ | \ \scrapelinkaction( \xpath ) \ | \ \downloadaction( \xpath)  \\ 
 & \ \ \ | \  
\gobackinstruction  \ | \  \extracturlaction \ | \ 
\sendkeysinstruction( \xpath, \conststring ) \ | \ \senddatainstruction( \xpath, \inputvaluepath ) \\ 
\xpath & \grammareq  \epsilon \ | \ \xpath/\predicate[i] \ | \ \xpath\sslash\predicate[i] \ \ \ \ \ \ \ \ \ \ \ \ \ \ \ 
\inputvaluepath  \grammareq  \inputvar \ | \ \inputvaluepath[\emph{key}] \ | \ \inputvaluepath[i] 
\end{array}
\]
Note that, different from statement $\statement$ in Figure~\ref{fig:dsl}, an action $\action$ is loop-free and uses \emph{concrete} selectors $\xpath$ and value paths $\inputvaluepath$.

We further illustrate how action tracking works using the following two simple rules (which include an environment in the output). Other rules are fairly similar. 
\[\footnotesize
\begin{array}{c}
\irule{
\evalprogram{\mystate, \env}{ \statement_1 }{ \actiontrace', \mystate', \env'} \ \ \ \ \ \ 
\evalprogram{\mystate', \env'}{ \statement_2, \mydots, \statement_m }{ \actiontrace\doubleprime, \mystate\doubleprime, \env\doubleprime }
}{
\evalprogram{\mystate, \env}{ \statement_1; \mydots; \statement_m }{ \actiontrace' \traceconcat \actiontrace\doubleprime, \mystate\doubleprime, \env\doubleprime }
} \ \ (\textsc{Seq})
\\ \\ 
\irule{
\evaldomnode{\env }{\domnode}{\xpath} \ \ \ \ \ \ 
\boxed{\mystate' = \texttt{Perform\_Click}(\xpath, \mystate)}
}{
\evalprogram{\mystate, \env}{\clickinstruction(\domnode)}{ [ \clickaction(\xpath) ], \mystate', \env }
} \ \ (\textsc{Click})
\end{array}
\]
The \textsc{Seq} rule is standard in that it executes $\statement_i$'s in sequence; however, note that it concatenates the action traces $\actiontrace'$ and $\actiontrace\doubleprime$ in the output. 
The actual action tracking takes place in the base rules, such as \textsc{Click}, but there is an issue: \textsc{Click} \emph{actually} performs the operation in the browser. This is problematic because during synthesis, candidate programs might have undesired side-effects (e.g., clicking a button that deletes the database), which prevents us from actually running them. 

This motivates \emph{our second key idea}: we \emph{simulate} the actual semantics without actually running $\program$, in particular, by simulating $\program$'s DOM transition using a trace $\statetrace$ of DOMs.\footnote{We can obtain this DOM trace by recording intermediate DOMs in tandem while recording the user-demonstrated actions.}
We illustrate how this simulation works still on \textsc{Seq} and \textsc{Click}. 
\[\footnotesize
\begin{array}{c}
\irule{
\evalprogram{\statetrace, \env}{ \statement_1 }{ \actiontrace', \statetrace', \env'} \ \ \ \ \ \ 
\evalprogram{\statetrace', \env'}{ \statement_2, \mydots, \statement_m }{ \actiontrace\doubleprime, \statetrace\doubleprime, \env\doubleprime }
}{
\evalprogram{\statetrace, \env}{ \statement_1; \mydots; \statement_m }{ \actiontrace' \traceconcat \actiontrace\doubleprime, \statetrace\doubleprime, \env\doubleprime }
} \ \ \ \ (\textsc{Seq})
\\ \\ 
\irule{
\statetrace = [ \mystate_1, \mydots, \mystate_m] \ \ \ \ \ \ 
\evaldomnode{\env }{\domnode}{\xpath} \ \ \ \ \ \ 
\boxed{\statetrace' = [ \mystate_2, \mydots, \mystate_m]}
}{
\evalprogram{\statetrace, \env}{\clickinstruction(\domnode)}{ [ \clickaction(\xpath) ], \statetrace', \env }
} \ \ \ \ (\textsc{Click})
\end{array}
\]
Here, instead of tracking the resulting DOM $\mystate'$, \textsc{Click} tracks the resulting DOM \emph{trace} $\statetrace'$; the intuition is that $\statetrace'$ contains DOMs that \emph{future actions} will be executed upon (instead of only the \emph{next immediate action}). 
The transition from $\statetrace$ to $\statetrace'$ is ``angelic'' in that \textsc{Click} \emph{always} transitions to the next DOM (by removing $\mystate_1$ from $\statetrace$), without actually doing the click on $\mystate_1$. But, what if performing the click on $\mystate_1$ does not yield $\mystate_2$? This is indeed possible, especially given that our synthesis algorithm often explores many wrong programs. However, if the resulting action trace $\actiontrace'$ matches the user-provided trace $\actiontrace$ (i.e., the specification), we know that every DOM transition must be genuine, because that is what we had recorded from the user demonstration (assuming deterministic replay). In other words, evaluating ``the right'' program that corresponds to $\statetrace$ guarantees to yield $\actiontrace'$ that matches the specification $\actiontrace$.

\newcounter{mycounter}
\begin{figure}[!t]
\scriptsize
\arraycolsep=1pt\def\arraystretch{1}
\begin{centermath}
\centering
\begin{array}{c}

\hspace{-210pt}
(\textsc{Eval})

\\[-8pt]

\irule{
\begin{array}{c}
\evalprogram{\statetrace, \{ \inputvar \mapsto \inputdata \} }{\program}{ \actiontrace', \statetrace', \env' }

\end{array}
}{
\evalprogramtop{\statetrace,  \inputdata  }{\program}{ \actiontrace' }
} 

\\ \\

\hspace{-210pt}
(\textsc{Term})

\\[-8pt]

\irule{
\begin{array}{c}
\statetrace = \emptylist
\end{array}
}{
\evalprogram{\statetrace, \env}{\program}{ \emptylist, \emptylist, \env }
} 

\\ \\ 

\hspace{-215pt}
(\textsc{Seq})

\\[-8pt]

\irule{
\begin{array}{c}
\evalprogram{
\statetrace, \env
}{
\block_{1}
}{
\actiontrace', \statetrace', \env' 
} \ \ \ \ 
\evalprogram{
\statetrace', \env'
}{
\block_{2}; \mydots; \block_m 
}{
\actiontrace\doubleprime, \statetrace\doubleprime,  \env\doubleprime
}
\\ 
\end{array}
}{
\evalprogram{ \statetrace, \env}{\block_1; \mydots; \block_m}{ 
\actiontrace' \actiontraceconcat \actiontrace\doubleprime, \statetrace\doubleprime,  \env\doubleprime}
}

\\ \\ 

\hspace{-208pt}
(\textsc{Click})

\\[-7pt]

\irule{
\begin{array}{c}
\statetrace = [ \mystate_1, \mydots, \mystate_m ] \ \ \ \ 
\evaldomnode{\env}{\domnode}{\xpath} 
\end{array}
}{
\evalprogram{
\statetrace,  \env
}{
\clickinstruction(\domnode)
}{
[ \clickaction(\xpath)  ], [ \mystate_2, \mydots, \mystate_m ], \env
}
} 

\\ \\ 

\hspace{-191pt}
(\textsc{EnterData})

\\ \\[-5pt]

\irule{
\begin{array}{c}
\statetrace = [ \mystate_1, \mydots, \mystate_m ] \ \ \ \ 
\evaldomnode{\env}{\domnode}{\xpath} \ \ \ \ 
\evalinputvalue{\env}{\inputvalue}{\inputvaluepath} 
\end{array}
}{
\evalprogram{ \statetrace, \env}{\senddatainstruction(\domnode, \inputvalue)}{ [ \senddataaction(\xpath, \inputvaluepath)  ], [\mystate_2, \mydots, \mystate_m], \env }
}

\\ \\ 

\hspace{-208pt}
(\textsc{S-Init})

\\[-8pt]

\begin{array}{c}

\irule{
\begin{array}{c}
\evalprogram{
\statetrace, \env
}{
\foreachloop{\domnodevar}{\domnodes_{\geq 1}}{ \program } 
}{
\actiontrace', \statetrace', \env' 
}
\end{array}
}{
\evalprogram{
\statetrace, \env
}{
\foreachloop{\domnodevar}{\domnodes}{ \program } 
}{
\actiontrace', \statetrace', \env' 
}
}

\\ \\ 

\hspace{-204pt}
(\textsc{S-Cont})

\\[-7pt]

\irule{
\begin{array}{c}
\statetrace = [ \mystate_1, \mydots, \mystate_m ] \ \ \ \ 
\evaldomnode{\env}{\domnodes_{\geq i}}{\xpath \listconcat \domnodes_{\geq i+1}} \ \ \ \ 
\valid( \xpath, \mystate_1 ) \\ 
\evalprogram{
\statetrace, \env[ \domnodevar \mapsto \xpath ]
}{
\program; \foreachloop{\domnodevar}{\domnodes_{\geq i+1}}{ \program } 
}{
\actiontrace', \statetrace', \env'
}
\end{array}
}{
\evalprogram{
\statetrace, \env
}{
\foreachloop{\domnodevar}{\domnodes_{\geq i}}{ \program } 
}{
\actiontrace', \statetrace', \env' 
}
}

\\ \\ 

\hspace{-206pt}
(\textsc{S-Term})

\\[-8pt]

\irule{
\begin{array}{c}
\statetrace = [ \mystate_1, \mydots, \mystate_m ] \ \ \ \ 
\evaldomnode{\env}{\domnodes_{\geq i}}{\xpath \listconcat \domnodes_{\geq i+1}} \ \ \ \ 
\neg\valid( \xpath, \mystate_1 ) 
\end{array}
}{
\evalprogram{
\statetrace, \env
}{
\foreachloop{\domnodevar}{\domnodes_{\geq i}}{ \program } 
}{
\emptylist, \statetrace, \env
}
}

\\ \\ 

\hspace{-202pt}
(\textsc{VP-Loop})

\\ \\[-14pt]

\irule{
\begin{array}{c}
\evalinputvalues{\env}{\inputvalues}{ [ \inputvaluepath_1, \mydots, \inputvaluepath_m ] } \ \ \ \ 
\statetrace_0 = \statetrace \ \ \ \ 
\env_0 = \env \\ 
\evalprogram{\statetrace_{i-1}, \env_{i-1}[\inputvaluevar \mapsto \inputvaluepath_i]  }{\program}{\actiontrace_i, \statetrace_i, \env_i} \ \ \ \ 
1 \leq i \leq m 
\end{array}
}{
\evalprogram{
\statetrace, \env}{
\foreachloop{\inputpathvar}{\inputvalues}{\program} 
}{
\actiontrace_1  \traceconcat \hspace{-1pt} \cdot \hspace{-1pt} \cdot \hspace{1pt} \traceconcat \actiontrace_m, 
\statetrace_m, 
\env_m
}
}

\\ \\ 

\hspace{-193pt}
(\textsc{While-Init})

\\ \\[-6pt]

\irule{
\begin{array}{c}
\evalprogram{
\statetrace, \env}{
\program; 
\
\myif{\valid(\domnode)}{
\big\{ 
\clickinstruction(\domnode); \while{\{ \program; \clickstatement(\domnode) \} }
\big\}}
}{
\actiontrace', \statetrace', \env' 
}
\end{array}
}{
\evalprogram{
\statetrace,  \env}{
\while{\{ \program; \clickstatement(\domnode)
\} }
}{
\actiontrace', \statetrace', \env' 
}
}

\\ \\ 

\hspace{-189pt}
(\textsc{While-Cont})

\\ \\[-5pt]

\irule{
\begin{array}{c}
\statetrace = [ \mystate_1, \mydots, \mystate_m ] \ \ \ \ 
\evaldomnode{\env}{\domnode}{\xpath} \ \ \ \ 
\valid(\xpath, \mystate_1) \ \ \ \ 
\evalprogram{\statetrace, \env}{\program}{\actiontrace', \statetrace', \env'}
\end{array}
}{
\evalprogram{
\statetrace,  \env}{
\myif{\valid(\domnode)}{\program}
}{
\actiontrace', \statetrace', \env'
}
} 

\\ \\ 

\hspace{-190pt}
(\textsc{While-Term})

\\ 

\irule{
\begin{array}{c}
\statetrace = [ \mystate_1, \mydots, \mystate_m ] \ \ \ \ 
\evaldomnode{\env}{\domnode}{\xpath} \ \ \ \ 
\neg\valid(\xpath, \mystate_1) 
\end{array}
}{
\evalprogram{
\statetrace,  \env}{
\myif{\valid(\domnode)}{\program}
}{
\emptylist, \statetrace, \env
}
} 
\end{array}

\end{array}
\end{centermath}
\vspace{-7pt}
\caption{Trace semantics of our web RPA language. 
}
\vspace{-12pt}
\label{fig:semantics}
\end{figure}

\textbf{\em Our trace semantics.}
Let us now explain our simulation semantics in detail. 
Our top-level judgment takes the form: 
\vspace{-3pt}
\[
\begin{array}{c}
\evalprogramtop{ \statetrace, \inputdata }{\program}{ \actiontrace' }
\end{array}
\vspace{-2pt}
\]
where $\actiontrace'$ is the action trace produced by $\program$, and $\statetrace$ is used to guide the simulated execution (as we also briefly discussed earlier). 
Our key rule is of the form: 
\vspace{-2pt}
\[
\begin{array}{c}
\evalprogram{\statetrace, \env}{\program}{\actiontrace', \statetrace', \env'}
\end{array}
\]
which intuitively states: 
\begin{myindent}
\vspace{2pt}
\noindent
\em 
Given DOM trace $\statetrace$ and environment $\env$, $\program$ would execute actions in $\actiontrace'$, yielding environment $\env'$ and DOM trace $\statetrace'$ (containing DOMs future actions will be executed upon). 
\vspace{2pt}
\end{myindent}
\noindent

\begin{figure}[!t]
\scriptsize
\arraycolsep=1pt\def\arraystretch{1}
\begin{centermath}
\centering
\begin{array}{c}

\begin{array}{c}

\begin{array}{lc}
(1) & 
\irule{
\begin{array}{c}
\end{array}
}{
\evaldomnode{\env}{\epsilon}{\epsilon}
} 

\\ \\ 

(2) & 
\irule{
\begin{array}{c}
\end{array}
}{
\evaldomnode{\env}{\domnodevar}{\env[\domnodevar]}
} 

\\ \\ 

(3) & 
\irule{
\begin{array}{c}
\evaldomnode{\env}{\domnode}{\xpath}
\end{array}
}{
\evaldomnode{\env}{
\domnode/\predicate[i]
}{\xpath/\xpathpredicate[i]}
}

\\ \\ 

(4)  \ \ & 
\irule{
\begin{array}{c}
\evaldomnode{\env}{\domnode}{\xpath} 
\end{array}
}{
\evaldomnode{\env}{
\domnode\sslash\predicate[i]
}{\xpath\sslash\xpathpredicate[i]}
} 
\end{array}
\ \ \ \ 
\begin{array}{lc}

(5) & 
\irule{
\begin{array}{c}
\end{array}
}{
\evalinputvalue{\env}{\inputvar}{\inputvar}
}

\\ \\ 

(6) & 
\irule{
\begin{array}{c}
\end{array}
}{
\evalinputvalue{\env}{\inputvaluevar}{\env[\inputvaluevar]}
}

\\ \\ 

(7) \ \ & 
\irule{
\begin{array}{c}
\evalinputvalue{\env}{\inputvalue}{\inputvaluepath}
\end{array}
}{
\evalinputvalue{\env}{\inputvalue[\emph{key}]}{\inputvaluepath[\emph{key}]}
}

\\ \\ 

(8) & 
\irule{
\begin{array}{c}
\evalinputvalue{\env}{\inputvalue}{\inputvaluepath}
\end{array}
}{
\evalinputvalue{\env}{\inputvalue[i]}{\inputvaluepath[i]}
}
\end{array}
\end{array}

\\ \\ 

\begin{array}{lc}
(9) \ & 
\irule{
\begin{array}{c}
\evaldomnode{\env}{\domnode}{\xpath} \ \ \ \ 
\evaldomnode{\env}{\domnode}{\xpath} \\
\end{array}
}{
\evaldomnodes{\env}{\domchildren( \domnode, \predicate)_{\geq i}}{\xpath\slash\xpathpredicate[i] \listconcat \domchildren( \xpath, \xpathpredicate)_{\geq i+1} }
}

\\ \\ 

(10) & 
\irule{
\begin{array}{c}
\evaldomnode{\env}{\domnode}{\xpath} \\
\end{array}
}{
\evaldomnodes{\env}{\domdescendants( \domnode, \predicate)_{\geq i}}{\xpath\sslash\xpathpredicate[i] \listconcat \domdescendants( \xpath, \xpathpredicate)_{\geq i+1} }
}

\\ \\ 

(11) & 
\irule{
\begin{array}{c}
\evalinputvalue{\env}{\inputvalue}{\inputvaluepath} \ \ \ \ 
\emph{arr} =  \emph{GetArray} ( \env[\inputvar], \inputvaluepath  )
\end{array}
}{
\evalinputvalues{\env}{\inputchildren(\inputvalue)}{ \big[ \inputvaluepath[1], \mydots, \inputvaluepath[|\emph{arr}|] \big] }
}
\end{array}

\end{array}
\end{centermath}
\vspace{-5pt}
\caption{Auxiliary rules for our trace semantics.}
\label{fig:aux-semantics}
\vspace{-10pt}
\end{figure}

\begin{figure*}
\scriptsize
\begin{centermath}
\hspace{-15pt}
\begin{array}{c}

\begin{array}{c}
\irule{

\begin{array}{c}
\irule{
\evaldomnode{ \{ \inputvar \mapsto \bot, \domnodevar \mapsto \sslash\emph{a}[2] \} }{ \domnodevar }{ 
\sslash\emph{a}[2] 
}
}{
\evalprogram{ [ \mystate_2 ], \{ \inputvar \mapsto \bot, \domnodevar \mapsto \sslash\emph{a}[2] \} }{\clickinstruction(\domnodevar)}{
[
\clickaction(\sslash\emph{a}[2]) 
], 
\emptylist, 
\{ \inputvar \mapsto \bot, \domnodevar \mapsto \sslash\emph{a}[2] \}
}
} (\textsc{Click}) 
\ \ \ \ \ \ 
\irule{
}{
\evalprogram{\emptylist, \mydots }{ \mydots }{ \emptylist, \emptylist, \{ \inputvar \mapsto \bot, \domnodevar \mapsto \sslash\emph{a}[2] \}}
}(\textsc{Term})
\end{array}

}{

\evalprogram{ [ \mystate_2 ], \{ \inputvar \mapsto \bot, \domnodevar \mapsto \sslash\emph{a}[2] \}}{
\clickinstruction(\domnodevar); 
\foreachloop{\domnodevar}{  \domdescendants(\epsilon, \emph{a})_{\geq 3}  }{ \{ \clickinstruction(\domnodevar) \}}
}{
[
\clickaction(\sslash\emph{a}[2]) 
], 
\emptylist, 
\{ \inputvar \mapsto \bot, \domnodevar \mapsto \sslash\emph{a}[2] \}
}

}(\textsc{Seq})
\end{array}
\\ \\ \\ 
\begin{array}{c}
\irule{
\evaldomnodes{\{  \domnodevar \mapsto \sslash\emph{a}[1] \}}{\domdescendants(\epsilon, \emph{a})_{\geq 2}}{\sslash\emph{a}[2]\listconcat\domdescendants(\epsilon, \emph{a})_{\geq 3}}
\ \ \ \ \ \ \ \ \ 
\emph{valid}(\sslash\emph{a}[2], \mystate_2)
\ \ \ \ \ \ \ \ \ 
\fbox{See above.}
}{

\evalprogram{ [ \mystate_2 ], \{ \inputvar \mapsto \bot, \domnodevar \mapsto \sslash\emph{a}[1] \}}{
\foreachloop{\domnodevar}{  \domdescendants(\epsilon, \emph{a})_{\geq 2}  }{ \{ \clickinstruction(\domnodevar) \}}
}{
[
\clickaction(\sslash\emph{a}[2]) 
], 
\emptylist, 
\{ \inputvar \mapsto \bot, \domnodevar \mapsto \sslash\emph{a}[2] \}
}
}
(\textsc{S-Cont})
\end{array}
\\ \\ \\
\begin{array}{c}
\irule{

\begin{array}{c}
\irule{
\evaldomnode{ \{ \inputvar \mapsto \bot, \domnodevar \mapsto \sslash\emph{a}[1] \} }{ \domnodevar }{ \sslash\emph{a}[1] }
}{
[ \mystate_1, \mystate_2 ], 
\evalprogram{ \{ \inputvar \mapsto \bot, \domnodevar \mapsto \sslash\emph{a}[1] \} }{ \clickinstruction(\domnodevar) }{ [ \clickaction(\sslash\emph{a}[1]) ], [ \mystate_2 ], \{  \inputvar \mapsto \bot, \domnodevar \mapsto \sslash\emph{a}[1] \}  }
}
(\textsc{Click})
\ \ \ \ \ \ \ \ \ \ \ \ 
\end{array}
\fbox{See above.}
}{
\evalprogram{ [ \mystate_1, \mystate_2 ], \{ \inputvar \mapsto \bot, \domnodevar \mapsto \sslash\emph{a}[1] \}}{
\clickinstruction(\domnodevar); 
\foreachloop{\domnodevar}{  \domdescendants(\epsilon, \emph{a} )_{\geq 2}  }{ \{ \clickinstruction(\domnodevar) \}}
}{
[
\clickaction(\sslash\emph{a}[1]), 
\clickaction(\sslash\emph{a}[2]) 
], 
\emptylist, 
\{ \inputvar \mapsto \bot, \domnodevar \mapsto \sslash\emph{a}[2] \}
}
}
(\textsc{Seq})
\end{array}
\\ \\ \\ 
\begin{array}{c}
\irule{

\irule{

\irule{

\evaldomnode{ \{ \inputvar \mapsto \bot \} }{\domdescendants(\epsilon, \emph{a} )_{\geq 1} }{ \sslash\emph{a}[1] \listconcat \domdescendants(\epsilon, \emph{a})_{\geq 2} }
\ \ \ \ \ \ 
\emph{valid}(\sslash\emph{a}[1], \mystate_1) 
\ \ \ \ \ \ 
\fbox{See above.}

}{
\evalprogram{ [ \mystate_1, \mystate_2 ], \{ \inputvar \mapsto \bot \}}{
\foreachloop{\domnodevar}{  \domdescendants(\epsilon, \emph{a})_{\geq 1}  }{ \{ \clickinstruction(\domnodevar) \}}
}{
[
\clickaction(\sslash\emph{a}[1]), 
\clickaction(\sslash\emph{a}[2]) 
], 
\emptylist, 
\{ \inputvar \mapsto \bot, \domnodevar \mapsto \sslash\emph{a}[2] \}
}
} 
(\textsc{S-Cont})

}{
\evalprogram{ [ \mystate_1, \mystate_2 ], \{ \inputvar \mapsto \bot \}}{
\foreachloop{\domnodevar}{  \domdescendants(\epsilon, \emph{a})  }{ \{ \clickinstruction(\domnodevar) \}}
}{
[
\clickaction(\sslash\emph{a}[1]), 
\clickaction(\sslash\emph{a}[2]) 
], 
\emptylist, 
\{ \inputvar \mapsto \bot, \domnodevar \mapsto \sslash\emph{a}[2] \}
}
}
(\textsc{S-Init})
}{
\evalprogramtop{ [ \mystate_1, \mystate_2 ], \bot }{
\foreachloop{\domnodevar}{ \domdescendants(\epsilon, \emph{a}) }{ \{ \clickinstruction(\domnodevar) \}}
}{
[ 
\clickaction(\sslash\emph{a}[1]), 
\clickaction(\sslash\emph{a}[2])
]
}
}
(\textsc{Eval})
\end{array}
\end{array}
\end{centermath}
\vspace{-5pt}
\caption{A derivation for the program in Example~\ref{ex:semantics} using our trace semantics.}
\vspace{-10pt}
\label{fig:semantics-derivation}
\end{figure*}

\textbf{\em Evaluating programs.}
The \textsc{Term} rule states that, if the input DOM trace is empty (\ie, there is no DOM to execute $\program$ upon), then we terminate the entire execution. 
Otherwise, we evaluate the statements sequentially using the \textsc{Seq} rule.

\textbf{\em Evaluating loop-free statements.}
Figure~\ref{fig:semantics} gives two example rules; the other rules are very similar. 
The \textsc{Click} rule first evaluates $\domnode$ to obtain a concrete selector $\xpath$ and then produces a $\clickaction$ action.
The \textsc{EnterData} rule is similar, except that it also evaluates the value path expression $\inputvalue$. As we can see, these rules form the base cases of our semantics.

\textbf{\em Evaluating loopy statements.}
The rest of the rules from Figure~\ref{fig:semantics} deal with loops. 
Amongst the first three rules that handle selector loops, the most interesting one perhaps is \textsc{S-Cont}: it unrolls the loop once if the first selector $\xpath$ refers to a DOM node that exists in  $\mystate_1$ (checked by \emph{valid}). 
This is another example for how we use DOMs to guide the simulated execution: we use DOMs to handle branches in loops. 
If $\xpath$ exists in $\mystate_1$ (e.g., the next element to be scraped exists), we bind $\domnodevar$ to $\xpath$ and execute the loop body $\program$. 
Note that \textsc{S-Cont} unrolls loops lazily. This is because many websites load more DOM nodes while scrolling down a page: we cannot eagerly fetch all DOM nodes at the beginning; instead, we have to keep executing until all nodes are loaded. 
The next rule, \textsc{VP-Loop}, handles value path loops. 
It is eager and it iterates over all value paths in $\inputvalues$. 
The last three rules handle while loops. A key distinction here is the termination condition: while loops are \emph{click-terminated}. That is, if the selector in the last $\clickstatement$ is not valid, it terminates. 
As mentioned earlier, this is mainly used to handle pagination using ``next page''. 
Note that, though not explicitly being defined, we use a standard \textbf{\texttt{if}} construct in our rules to help formalize the semantics.

\textbf{\em Auxiliary rules.}
Figure~\ref{fig:aux-semantics} presents the auxiliary rules for evaluating symbolic selectors and value paths. 
They are fairly straightforward. 
For instance, rules (1)-(4) handle selector expressions that may contain variables, by basically replacing variables with concrete values. Rules (5)-(8) are conceptually the same except that they are for symbolic value paths. Rules (9)-(11) evaluate selectors expressions. 

\begin{example}
Consider the following program $\program$, which is an extremely simplified version of $\program_1$ from Figure~\ref{fig:demo}(c). 
\[
\small
\hspace{-10pt}
\begin{array}{l}
\begin{array}{l}
\foreachloop{ \domnodevar }{\domdescendants(\epsilon, \emph{a})}{ 
\{ \clickinstruction( \domnodevar ) \}
} 
\end{array}
\end{array}
\vspace{-2pt}
\]
Here, $\program$ performs a $\clickaction$ (using variable $\domnodevar$) in a selectors loop. 
For simplicity, let us consider a DOM trace $ \statetrace = [ \mystate_1, \mystate_2 ]$. Let us also assume $\statetrace$ indeed corresponds to $\program$; that is, the $i$-th click in $\program$ executed on DOM $\mystate_i$ transitions the page to $\mystate_{i+1}$.

We illustrate our semantics on $\program$; 
Figure~\ref{fig:semantics-derivation} shows its derivation. 
First of all, \textsc{Eval} returns two actions that are executed in the first two iterations of $\program$, since $\statetrace$ has two DOMs. The \emph{valid} checks in \textsc{S-Cont} are used to guide our simulated execution. For $\program$, these checks all pass, as $\program$ indeed produced $\statetrace$. 
However, consider the following $\program'$: 
\hspace{-10pt}
\[\small
\begin{array}{l}
\begin{array}{l}
\foreachloop{ \domnodevar }{\domdescendants(\epsilon, \emph{a})}{ 
\{ \clickinstruction( \domnodevar \slash \emph{b} ) \}
} 
\end{array}
\end{array}
\]
For $\program'$, the checks may not pass as ``$\sslash\emph{a}[1]/\emph{b}$'' might not refer to a valid node in DOM $\mystate_1$. In that case, we will invoke the \textsc{S-Term} rule, eventually producing a shorter action trace. 
\label{ex:semantics}
\end{example}

%% file: problem.tex
\section{Web RPA Program Synthesis Problem}
\label{sec:problem}

In this section, we formulate our program synthesis problem. 

\begin{definition}
\hspace{-3pt}(Satisfaction).
Given input data $\inputdata$, an action trace $\actiontrace$ and a DOM trace $\statetrace$, 
a web RPA program $\program$ \emph{satisfies} $\actiontrace$, if we have (1) $\evalprogramtop{\statetrace, \inputdata }{\program}{\actiontrace'}$ and (2) $\actiontrace$ is \emph{consistent} with a \emph{prefix} of $\actiontrace'$ with respect to $\statetrace$. In other words, $\program$ can \emph{reproduce} $\actiontrace$. 
\label{def:sat}
\end{definition}

Definition~\ref{def:sat} requires checking consistency between two traces of actions. 
To do this, we first define two actions $\action_1$ and $\action_2$ to be consistent, given a DOM $\mystate$, if $\action_1$ and $\action_2$ are of the same type and their arguments match. 
Note that two XPath arguments match each other, if they refer to the same DOM node on $\mystate$. Then, two action traces $\actiontrace_1$ and $\actiontrace_2$ are consistent, given a DOM trace $\statetrace$, if the $i$-th action in $\actiontrace_1$ is consistent with the $i$-th action in $\actiontrace_2$ given the $i$-th DOM in $\statetrace$. 

The reason that condition (2) uses ``prefix'', instead of requiring $\actiontrace$ to be consistent with $\actiontrace'$ exactly, is because $\actiontrace$ is in general an \emph{incomplete} trace. 
That is, $\actiontrace$ may be a prefix of the entire action trace of $\program$. In other words, our trace semantics might produce a longer action trace than the demonstration.

\begin{definition}
\hspace{-3pt}(Generalization).
Given input data $\inputdata$, an action trace $\actiontrace$ and a DOM trace $\statetrace$, 
a web RPA program $\program$ \emph{generalizes} $\actiontrace$, 
if we have (1) $\evalprogramtop{\statetrace, \inputdata }{\program}{\actiontrace'}$ and (2) $\actiontrace$ is consistent with a \emph{strict} prefix of $\actiontrace'$ given $\statetrace$. In other words, $\program$ not only reproduces $\actiontrace$ but also executes more actions after $\actiontrace$. 
\label{def:generalization}
\end{definition}

Definition~\ref{def:generalization} requires ``strict prefix'', as our goal is to predict unseen actions beyond only reproducing those observed. 

\begin{definition}
\hspace{-3pt}(Web RPA Program Synthesis Problem).
Given input data $\inputdata$, an action trace $\actiontrace = [ \action_1, \mydots, \action_m]$ and a DOM trace $\statetrace = [\mystate_1, \mydots, \mystate_{m+1} ]$, 
find a web RPA program $\program$ that \emph{generalizes} $\actiontrace$, given $\inputdata$ and $\statetrace$. 
\label{def:problem}
\end{definition}

\vspace{-4pt}
Intuitively, Definition~\ref{def:problem} takes as input $\actiontrace$ and $\statetrace$ where $\action_i$ is an action performed on $\mystate_i$, and it looks for a program $\program$ that can be used to \emph{predict} an action $\action_{m+1}$ that might be performed on $\mystate_{m+1}$. 
In general, we require $\statetrace$ be longer than $\actiontrace$; otherwise, we are not able find a program that generalizes. In practice, we require $\statetrace$ have one more element than $\actiontrace$, because we can obtain the latest DOM without knowing the user's next action on it. Also note that, we may have multiple programs that generalize $\actiontrace$; therefore, we aim to synthesize a \emph{smallest} program in size.

%% file: algorithm.tex
\section{Web RPA Program Synthesis Algorithm}
\label{sec:algorithm}

\subsection{Top-Level Rewrite-Based Synthesis Algorithm}

Algorithm~\ref{alg:top} shows the top-level synthesis algorithm. 
Our key idea is to \emph{iteratively rewrite} the action trace $\actiontrace$ into a program that generalizes $\actiontrace$ given DOM trace $\statetrace$ and input data $\inputdata$. The algorithm is not destructive and maintains intermediate rewrites; it heuristically picks a ``best'' program at the end. 

Algorithm~\ref{alg:top} maintains a worklist of tuples $(\program, \vec{\actiontrace}, \vec{\statetrace})$, where $\program = \statement_1; \mydots; \statement_l$ is a program rewritten from the input trace $\actiontrace$.  $\vec{\actiontrace} = [\actiontrace_1, \mydots, \actiontrace_l]$ is a list of action traces, and $\vec{\statetrace} = [ \statetrace_1, \mydots, \statetrace_l ]$ is a list of DOM traces. We maintain the following invariant:
\vspace{-9pt}
\[\small 
\begin{array}{c}
\mathcal{I}_1: \ \ \ \ 
\actiontrace_1 \traceconcat \cdot \hspace{-1pt} \cdot  \hspace{2pt} \traceconcat \actiontrace_l = A 
\text{ and } 
\statetrace_1 \traceconcat \cdot \hspace{-1pt} \cdot  \hspace{2pt} \traceconcat \statetrace_l = [\mystate_1, \mydots, \mystate_m] 
\end{array}
\vspace{-2pt}
\]
Essentially, $\mathcal{I}_1$ says $\vec{\actiontrace}$ is a partition of  $\actiontrace$ and $\vec{\statetrace}$ is a partition of the first $m$ DOMs in $\statetrace$. 
It is fairly easy to show $\mathcal{I}_1$ holds for $\program_0, \vec{\actiontrace}_0, \vec{\statetrace}_0$. 
The second invariant is:  
\vspace{-1pt}
\[\small 
\begin{array}{c}
\mathcal{I}_2: 
\ \ \ \ 
\forall i \in [ 1, l ], 
\statement_i \emph{ satisfies } \actiontrace_i \emph{ given } \inputdata \emph{ and } \statetrace_i
\end{array}
\vspace{-2pt}
\]
which says that each $\statement_i$ in $\program$ \emph{satisfies} the corresponding \emph{slice} $\actiontrace_i$. This is also trivially true for $\program_0, \vec{\actiontrace}_0, \vec{\statetrace}_0$, as every $\statement_i$ in $\program_0$ is a single loop-free statement. These invariants guarantee our rewrites always \emph{satisfy} the specification $\actiontrace$. 
Intuitively, this is because every statement $\statement_i$ in $\program$ satisfies each slice $\actiontrace_i$ in $\actiontrace$, thus the ``concatenation'' of all $\statement_i$'s, that is $\program$, would also satisfy the concatenation of all $\actiontrace_i$'s, which is $\actiontrace$.

The worklist algorithm maintains $\mathcal{I}_1$ and $\mathcal{I}_2$, too. It tracks a worklist $\worklist$ of programs that \emph{satisfy} $\actiontrace$ but only stores those \emph{generalizable} programs into $\setprograms$. 
In particular, the algorithm first removes a tuple $( \program, \vec{\actiontrace}, \vec{\statetrace} )$ from $\worklist$ (line 4).
It then checks if $\program$ generalizes $\actiontrace$; if so, it adds $\program$ into $\setprograms$ (line 5). The algorithm grows the worklist using our \emph{speculate-and-validate} method to rewrite $\program$ into more programs all of which maintain $\mathcal{I}_1$ and $\mathcal{I}_2$ (lines 6-7). Intuitively, given $\program = \statement_1; \mydots; \statement_l$, this rewrite process replaces \emph{a slice of statements} $\statement_i, \mydots, \statement_j$ in $\program$ with a loop statement $\statement'$ such that $\statement_1; \mydots; \statement_{i-1}; \statement'; \statement_{j+1}; \mydots; \statement_l$ also meets $\mathcal{I}_1$ and $\mathcal{I}_2$. 
Note that, because a statement in $\program$ might itself be loopy, we can generate nested loops (from the inside out).

\begin{figure}[!t]
\vspace{-10pt}
\begin{algorithm}[H] 
\footnotesize
\begin{algorithmic}[1]
\myprocedureindent{Synthesize}{$\actiontrace, \statetrace, \inputdata$}{-15pt}

\Statex\Input{$\actiontrace = [ \action_1, \mydots, \action_m ]$, $\statetrace = [ \mystate_1, \mydots, \mystate_{m+1} ]$, and input data $\inputdata$.}
\Statex\Output{a program $\program$ that generalizes $\actiontrace$ given $\statetrace$ and $\inputdata$.}

\State $\program_0 \assign \action_1; \mydots; \action_m; 
\ \ 
\vecactiontrace_0 \assign \big[ [ \action_1 ], \mydots, [ \action_m ] \big]; 
\ \ 
\vecstatetrace_0 \assign \big[ [ \mystate_1 ], \mydots, [ \mystate_m ] \big];$

\State $\worklist \assign \{ ( \program_0, \vec{\actiontrace}_0, \vec{\statetrace}_0 ) \}; \ \ 
\setprograms \assign \emptyset$;

\While{$\worklist \neq \emptyset$} 

\State $( \program, \vec{\actiontrace}, \vec{\statetrace} ) \assign \worklist.\textsf{remove}()$;

\If{$\program \emph{ generalizes } \actiontrace \emph{ given } \inputdata \emph{ and } \statetrace$} $\setprograms.\textsf{add}( \program )$; 
\EndIf

\State $\srewrites \assign \textsc{Speculate}(\program)$;

\State $\rewrites' \assign \textsc{Validate}(\srewrites, \program, \vecactiontrace, \vecstatetrace); \ \ 
\worklist \assign \worklist \setor \rewrites'$;

\EndWhile

\State \Return $\textsc{Rank}(\setprograms)$; 

\caption{Top-level synthesis algorithm.}
\label{alg:top}
\end{algorithmic}
\end{algorithm}
\vspace{-20pt}
\end{figure}

\textbf{\em Challenges.}
While conceptually simple, this idea is technically quite challenging to realize. 
A key challenge is that, it is in general quite hard to encode all patterns as rules, if we follow standard rewrite-based synthesis approaches~\cite{nandi2020synthesizing, willsey2021egg}: our DSL has multiple types of loops, a loop body may have multiple statements that may use loop variables in different ways, loops could be nested, \etc. There are too many cases.
Even if we can define these rules, it is not clear how efficient this rule-based approach is, given the trace may correspond to an arbitrarily complex program. 
Let us further illustrate this challenge using the following example. 

\begin{example}
Consider the following program $\program$, which is a simplified version of $\program_2$ in Figure~\ref{fig:demo}(c). It scrapes information from a list of items spanned across multiple pages. 
\[\small 
\begin{array}{l}
\begin{array}{l}
\while{} \\ 
\ \ \ \ \foreachloop{ \domnodevar }{\domdescendants(\epsilon, \emph{a})}{ }  \\ 
\ \ \ \ \ \ \ \ \scrapetextinstruction( \domnodevar ) \\
\ \ \ \ \ \ \ \ \scrapetextinstruction( \domnodevar\slash\emph{b}) \\
\ \ \ \ \clickstatement(\emph{c})
\end{array}
\end{array}
\]
\label{ex:rulebasedrewrite}
Suppose we are given the following action trace $\actiontrace$ for $\program$: 
\[\small
\begin{array}{lc}
[ \ 
\scrapetextaction(\sslash \emph{a}[1]), \scrapetextaction(\sslash \emph{a}[1]\slash\emph{b}), \mydots,  \\ 
\ \ \ 
\scrapetextaction(\sslash \emph{a}[20]), 
\scrapetextaction(\sslash \emph{a}[20]\slash\emph{b}), 
& \hspace{10pt} \emph{($\action_1, \mydots, \action_{40}$)}
\\ 
\ \ \ \clickstatement(\emph{c}), 
& \hspace{30pt} \emph{($\action_{41}$)} 
\\ 
\ \ \ \scrapetextaction(\sslash \emph{a}[1]), 
\scrapetextaction(\sslash \emph{a}[1]\slash\emph{b}), \mydots,  \\ 
\ \ \
\scrapetextaction(\sslash \emph{a}[9]), 
\scrapetextaction(\sslash \emph{a}[9]\slash\emph{b})
\ ] 
& \hspace{10pt} \emph{($\action_{42}, \mydots, \action_{59}$)}
\end{array}
\]
Here, $\action_1, \mydots, \action_{41}$ correspond to \emph{all} actions from the first iteration of the {\color{magenta}{\textbf{\texttt{while}}}} loop, including 40 actions from  {\color{magenta}{\textbf{\texttt{foreach}}}}. The remaining actions $\action_{42}, \mydots, \action_{59}$ correspond to a partial execution of the second iteration of {\color{magenta}{\textbf{\texttt{while}}}}. 
We also record a DOM trace $[\mystate_1, \mydots, \mystate_{60}]$ as well, where $\action_i$ is performed on $\mystate_i$ and $\mystate_{60}$ is the latest DOM.

In order to generate $\program$ from $\actiontrace$, the standard rewrite-based synthesis approaches~\cite{nandi2020synthesizing, willsey2021egg} apply a set of predefined \emph{sound} rewrite rules to rewrite $\actiontrace$ to $\program$. Conceptually, it would use these rules to essentially identify iteration boundaries and repetitive patterns, in order to eventually ``reroll'' the trace back to the desired program with loops. That is an enormous space which might contain only a few correct rewrites. 
\label{ex:challenge}
\end{example}

In this work, we take a different route which incorporates the ``guess-and-check'' idea into the overall rewrite process. Our approach does not generate true rewrites \emph{directly} using sound rewrite rules; instead, it first \emph{speculates} likely rewrites which are then \emph{validated} using our trace semantics.

\subsection{Speculation}

\begin{figure}[!t]
\vspace{-10pt}
\begin{minipage}{.98\linewidth}
\begin{algorithm}[H] 
\scriptsize

\begin{algorithmic}[1]
\vspace{-2pt}
\myprocedureindent{Speculate}{$\program$}{-15pt}

\Statex\Input{$\program = \statement_1; \mydots; \statement_l$.}

\Statex\Output{a set $\srewrites$ of speculative rewrites of the form $(\statement', \statement_i, \statement_j)$.}

\State $\srewrites \assign \emptyset$;

\vspace{1pt}

\ForAll{$i \leq p \leq j < q$ \emph{s.t.} $[ \statement_i, \mydots, \statement_p, \mydots, \statement_j, \mydots, \statement_q ] \sublisteq [ \statement_1, \mydots, \statement_l ]$ \& $j-i+1=q-p$}

\ForAll{$( \statement'_p, \domnodevar, \domnodes ) \in \textsc{Anti-Unify} ( \statement_{p}, \statement_{q} )$} 

\State $\xpath \assign \textsf{FirstSelector}(\domnodes)$;
\State $\setprograms' \assign \big\{  \statement'_i; \mydots; \statement'_p; \mydots; \statement'_j \  \big| \ \statement'_{k} \in \textsc{Parametrize}( \statement_k, \domnodevar, \xpath ), k  \in [i, j] \setminus \hspace{-2pt} \{p\} \big\}$;

\State $\statements' \assign \big\{ \foreachloop{\domnodevar}{\domnodes}{ \program' } \ | \ \program' \in \setprograms' \big\}$;

\State $\srewrites \assign \srewrites \cup \{ ( \statement', \statement_i, \statement_j ) \ | \ \statement' \in \statements'  \} $;

\EndFor
\EndFor

\vspace{1pt}

\ForAll{$i \leq p \leq j < q$ \emph{s.t.} $[ \statement_i, \mydots, \statement_p, \mydots, \statement_j, \mydots, \statement_q ] \sublisteq [ \statement_1, \mydots, \statement_l ]$ \& $j-i+1 =q-p$}

\ForAll{$( \statement'_p, \inputvaluevar, \inputvalues ) \in \textsc{Anti-Unify} ( \statement_{p}, \statement_{q} )$}

\State $\inputvaluepath \assign \textsf{FirstValuePath}(\inputvalues)$;

\State $\setprograms' \assign \big\{  \statement'_i; \mydots; \statement'_p; \mydots; \statement'_j  \ \big| \  \statement'_{k} \in \textsc{Parametrize}( \statement_l, \inputvaluevar, \inputvaluepath ), k  \in [i, j] \ \{p\} \big\}$;

\State $\statements' \assign \big\{ \foreachloop{\inputvaluevar}{\inputvalues}{ \program' } \ \big| \ \program' \in \setprograms' \big\}$;

\State $\srewrites \assign \srewrites \cup \{ ( \statement', \statement_i, \statement_j ) \ | \ \statement' \in \statements'  \} $;

\EndFor
\EndFor

\vspace{1pt}

\ForAll{$i < p < q$ \emph{s.t.} $[ \statement_i, \mydots, \statement_p, \dots, \statement_q ] \sublisteq [ \statement_1, \mydots, \statement_l ]$ \& $p-i+1=q-p$}

\If{$\statement_p = \statement_q = \clickstatement(\xpath)$}
\State $\statement' \assign \while{ \{ \statement_i, \mydots, \statement_p \} } $; \ \ $\srewrites \assign \srewrites\setor \{ ( \statement', \statement_i, \statement_j ) \}$;
\EndIf 

\EndFor

\vspace{1pt}

\State\Return $\srewrites$;

\caption{\textsc{Speculate} procedure.}
\label{alg:speculate}

\end{algorithmic}
\end{algorithm}
\vspace{-20pt}
\end{minipage}
\end{figure}

We first describe our speculation procedure; see Algorithm~\ref{alg:speculate}. 
It takes as input a program $\program = \statement_1; \mydots; \statement_l$
and returns a set $\srewrites$ of \emph{speculative rewrites}, or \emph{s-rewrites}, of the form $(\statement', \statement_i, \statement_j)$. Here,  $\statement'$ is a loop statement whose \emph{first} iteration corresponds to $\statement_i; \mydots; \statement_j$ from $\program$. That is, they yield the same trace (this is guaranteed by construction). 
However, an s-rewrite may not be a true rewrite: a true rewrite must have \emph{more than one iterations} exhibited in $\program$, but an s-rewrite is only guaranteed to have its first iteration exhibited in $\program$. 
Nevertheless, s-rewrites have a very nice property: they are much easier to generate, and they over-approximate the set of true rewrites. Our technique makes use of this property.

To generate s-rewrites that \emph{tightly over-approximate} the set of true rewrites, we follow existing rule-based approaches in prior work~\cite{nandi2020synthesizing, willsey2021egg}. However, our rules are designed to detect patterns \emph{partially}, instead of completely. 
A key step in our approach is to inspect two statements $\statement_p, \statement_q$ in $\program$ and generate a loop $\statement'$ \emph{such that}, $\statement_p,\statement_q$ correspond to the \emph{same} statement from $\statement'$ but $\statement_p$ ``comes'' from its first iteration and $\statement_q$ from its second. 
For example, lines 2-7 in Algorithm~\ref{alg:speculate} generate selector loops. It first enumerates all slices $\statement_i, \mydots, \statement_j$ in $\program$ assuming $\statement_i$ and $\statement_j$ correspond to the start and end of the first iteration (line 2). 
Then, it tries to  ``merge'' $\statement_p, \statement_q$ into a \emph{parametrized} statement $\statement'_p$ by calling \textsc{Anti-Unify} (line 3). 
Similarly, lines 8-13 handle value path loops.

\resetmycounter{mycounter}
\begin{figure}
\scriptsize
\arraycolsep=.5pt\def\arraystretch{1}
\begin{centermath}
\centering
\begin{array}{lc}
(1) \ \  & 
\irule{
\begin{array}{c}
\domnodevar \text{ fresh} \ \ \ \ 
\unifyvar{\domnodevar}{\xpath_1}{\xpath_2}{(\domnode, \domnodes)} 
\end{array}
}{
\unify{\clickinstruction( \xpath_1 )}{\clickinstruction( \xpath_2 ) }
{ ( \clickinstruction(\domnode), \domnodevar, \domnodes )}
}
\\ \\ 
(2) \ \ & 
\irule{
\begin{array}{c}
\domnodevar' \text{ fresh} \ \ \ \ 
\unifyvar{\domnodevar'}{\domnodes_1}{\domnodes_2}{ ( \domnodes, \domnodes' ) } \ \ \ \ 
\program_1,\program_2 \ \emph{alpha equivalent} \\ 
U =
(
\foreachloop{\domnodevar_1}{\domnodes}{\program_1}, 
\domnodevar', 
\domnodes'
) 
\end{array}
}{
\unify{\foreachloop{\domnodevar_1}{\domnodes_1}{\program_1}}
{\foreachloop{\domnodevar_2}{\domnodes_2}{\program_2}} 
{U}
}
\\ \\ 
(3) \ \ & 
\irule{
\begin{array}{c}
\inputvaluevar \text{ fresh} \ \ \ \ 
\inputvaluepath_1 = \inputvaluepath[1][o_1] \hspace{-2pt}\cdot\hspace{-2pt}\cdot [o_r] \ \ \ \ 
\inputvaluepath_2 = \inputvaluepath[2][o_1] \hspace{-2pt}\cdot\hspace{-2pt}\cdot [o_r]  \\ 
U = 
( \senddatastatement(\xpath, \inputvaluevar[o_1] \hspace{-2pt}\cdot\hspace{-2pt}\cdot [o_r]), \inputvaluevar, \inputchildren(\inputvaluepath) )
\end{array}
}{
\unify{\senddatastatement( \xpath, \inputvaluepath_1 )}{\senddatastatement( \xpath, \inputvaluepath_2 ) }
{U}
}
\\ \\ 
(4) \ \  & 
\irule{
\begin{array}{c}
\exists \xpath'_1 \hspace{-1pt}\in\hspace{-1pt} \emph{AlternativeSelectors}(\xpath_1) \hspace{-2pt}: \xpath'_1 = \domnode \big[ \domnodevar \mapsto \xpath\slash\predicate[1] \big] \\ 
\exists \xpath'_2 \hspace{-1pt}\in\hspace{-1pt} \emph{AlternativeSelectors}(\xpath_2) \hspace{-2pt}: \xpath'_2 = \domnode \big[ \domnodevar \mapsto \xpath\slash\predicate[2] \big]
\end{array}
}{
\unifyvar{\domnodevar}{\xpath_1}{\xpath_2}{ \big( \domnode, \domchildren( \xpath, \predicate ) \big)} 
}
\\ \\ 
(5) \ \  & 
\irule{
\begin{array}{c}
\unifyvar{\domnodevar}{\xpath_1}{\xpath_2}{(\domnode, \domnodes)} 
\end{array}
}{
\unifyvar{\domnodevar}{\domchildren( \xpath_1, \predicate )}{\domchildren( \xpath_2, \predicate )}
{
(
\domchildren( \domnode, \predicate ), \domnodes 
)
}
}
\end{array}
\end{centermath}
\vspace{-8pt}
\caption{\textsc{Anti-Unify} rules.}
\label{fig:anti-unify}
\vspace{-12pt}
\end{figure}

\textbf{\em Anti-unification.}
In the context of logic programming, anti-unification~\cite{baumgartner2017higher,baumgartner2014unranked} refers to the process of generating two terms $t_1$ and $t_2$ into a least general template $\tau$ for which there exists substitutions $\alpha_1$ and $\alpha_2$, such that $\tau(\alpha_1)=t_1$ and $\tau(\alpha_2)=t_2$. 
It has been used in prior work~\cite{sousa2021learning} to generate code fixes; in this work, we use anti-unification to synthesize loops. 
Figure~\ref{fig:anti-unify} gives some representative rules. 
In a nutshell, our procedure returns a set of tuples  $(\statement'_p, \domnodevar, \domnodes)$, where $\statement'_p$ is a more general statement using loop variable $\domnodevar$ in the target loop $\statement'$, and $\domnodes$ is the selectors that $\statement'$ loops over.

Let us take rule (1) as an example. Here, it anti-unifies two $\clickinstruction$ statements whose selectors differ at only one index in their XPath expressions. Specifically, it calls rule (4) that anti-unifies selectors $\xpath_1$ and $\xpath_2$ given a fresh variable $\domnodevar$. Intuitively, it looks for a general selector $\domnode$ that uses variable $\domnodevar$ such that $\domnode$ instantiates to $\xpath'_1$ and $\xpath'_2$, respectively. 
Note that rule (4) considers alternative selectors; this is necessary for inducing more general programs. Rule (4) also returns $\domchildren( \xpath, \predicate )$ which is the collection that the target loop statement $\statement'$ loops over. We have a very similar rule that anti-unifies $\xpath_1, \xpath_2$ and generates $\domdescendants( \xpath, \predicate )$, though it is not shown here. 

Rule (2) anti-unifies two selector loops by anti-unifying their respective collections $\domnodes_1$ and $\domnodes_2$, which is conditional on their loop bodies $\program_1, \program_2$ being alpha-equivalent. 
Rule (3) performs anti-unification for $\senddatastatement$ statements. 

\begin{example}
Consider the action trace $\actiontrace$ from Example~\ref{ex:challenge}. 
Line 2 of our \textsc{Speculate} procedure will consider all possible tuples $(i, p, j, q)$, where $i, j$ are the start and end of the first iteration of a loop to be generated. 
Consider $(1,1,2,3)$, which corresponds to the first iteration of the {\color{magenta}{\textbf{\texttt{foreach}}}} loop in $\program$. \textsc{Anti-Unify} at line 3 generates $(\scrapetextinstruction(\domnodevar), \domnodevar, \domdescendants(\epsilon, \emph{a}))$ from $\statement_p, \statement_q$, which are $\action_1, \action_3$  in this case. 
Here, $\scrapetextinstruction(\domnodevar)$ is the desired statement in the {\color{magenta}{\textbf{\texttt{foreach}}}} loop's body. Furthermore, it also gives the selectors expression, $\domdescendants(\epsilon, \emph{a})$, that the target {\color{magenta}{\textbf{\texttt{foreach}}}} loop iterates over. Yet, our \textsc{Anti-Unify} procedure does not generate the rest of the body.
\label{ex:anti-unify}
\vspace{-3pt}
\end{example}

\textbf{\em Parametrization}
\textsc{Anti-Unify} essentially creates a skeleton of the entire loop $\statement'$: it gives one statement in the loop body but we still need to construct the rest. 
This is exactly what Algorithm~\ref{alg:speculate} does at lines 4-7. In particular, it first obtains the binding $\domnodevar \mapsto \xpath$ in the first iteration. Then, given this binding, it uses the \textsc{Parametrize} procedure to construct the entire loop. 
Figure~\ref{fig:parametrize} presents some representative rules. 
For instance, rules (1) and (2) parametrize a $\clickinstruction$ statement. Rule (1) keeps the $\clickinstruction$ as is, since it is possible that a statement inside a loop does not use the variable. Rule (2) parametrizes the $\clickinstruction$ if the selector $\xpath'$ that variable $\domnodevar$ binds to is a prefix of some \emph{alternative selector} for the argument $\xpath$ in $\clickinstruction$. 
Rules (3) and (4) parametrize a selectors loop in a very similar way,  though it uses additional rules (5) and (6) to handle selectors.

\resetmycounter{mycounter}
\begin{figure}[!t]
\begin{centermath}
\centering
\scriptsize
\arraycolsep=.5pt\def\arraystretch{1}
\begin{array}{c}
(1) \  
\irule{
\statement = \clickstatement( \xpath )
}{
\parameterize{\domnodevar, \xpath'}{\clickstatement(\xpath)}
{ 
\statement
}
} 
\ \ \ \ 
(2) \ 
\irule{
\exists \xpath\doubleprime \in \emph{AlternativeSelectors}(\xpath) : \xpath\doubleprime = \xpath' / \xpath\tripleprime
}{
\parameterize{\domnodevar, \xpath'}{\clickstatement( \xpath )}
{ 
\clickstatement(\domnodevar/\xpath\tripleprime)
}
} 
\\ \\ 
(3) \ 
\irule{
\statement = \foreachloop{\domnodevar}{\domnodes}{\program}
}{
\parameterize{\domnodevar', \xpath}{
\foreachloop{\domnodevar}{\domnodes}{\program}
}{
\statement
}
} 
\\ \\ 
(4) \ 
\irule{
\begin{array}{c}
\parameterize{\domnodevar', \xpath}{\domnodes}{\domnodes'} 
\end{array}
}{
\parameterize{\domnodevar', \xpath}{
\foreachloop{\domnodevar}{\domnodes}{\program}
}{
\foreachloop{\domnodevar}{\domnodes'}{\program}
}
} 
\\ \\ 
(5) \ 
\irule{
\domnodes = \domchildren( \xpath, \predicate )
}{
\parameterize{\domnodevar, \xpath'}{
\domchildren( \xpath, \predicate )
}{
\domnodes
}
} 
\ \ \ \ 
(6) \ 
\irule{
\begin{array}{c}
\xpath = \xpath'/\xpath\doubleprime \ \ \ \ 
\domnodes = \domchildren( \domnodevar/\xpath\doubleprime, \predicate )
\end{array}
}{
\parameterize{\domnodevar, \xpath'}{
\domchildren( \xpath, \predicate )
}{
\domnodes
}
} 
\end{array}
\end{centermath}
\vspace{-5pt}
\caption{\textsc{Parametrize} rules.}
\vspace{-10pt}
\label{fig:parametrize}
\end{figure}

\begin{example}
Consider the output of \textsc{Anti-Unify}, namely, {\small $(\scrapetextinstruction(\domnodevar), \domnodevar, \domdescendants(\epsilon, \emph{a}))$}, in Example~\ref{ex:anti-unify}. 
Given this output, line 4 of Algorithm~\ref{alg:speculate} obtains the first selector $\xpath$ of $\domdescendants(\epsilon, \emph{a})$; that is, $\xpath = \sslash\emph{a}[1]$. 
Then, we parametrize each of the remaining statements within $[i, j]$---in our case, only $\action_2$. 
One statement given by \textsc{Parametrize} is  $\scrapetextinstruction(\domnodevar\slash\emph{b})$, which is the desired statement in $\program$. 
Therefore, $\statements'$ at line 6 of Algorithm~\ref{alg:speculate} includes the desired  {\color{magenta}{\textbf{\texttt{foreach}}}} loop. 
Finally, line 7 adds the following s-rewrite to $\srewrites$. 
\[\small 
\begin{array}{c}
\Big(
\begin{array}{l}
\foreachloop{\domnodevar}{\domdescendants(\epsilon, \emph{a})}{} \\ 
\ \ \ \ \scrapetextinstruction(\domnodevar)  \\ 
\ \ \ \ \scrapetextinstruction(\domnodevar\slash\emph{b})
\end{array}
,
\ \ \action_1,
\ \ \action_2 
\ \ \ 
\Big)
\end{array}
\]
While this loop corresponds to $\action_1, \mydots, \action_{40}$, our \textsc{Speculate} procedure only guarantees its first iteration corresponds to $\action_1, \action_2$. 
\label{ex:parametrize}
\vspace{-12pt}
\end{example}

\subsection{Validation}
As we can see, s-rewrites are fairly easy to generate but may be spurious. 
That is, they might not rewrite beyond the first iteration. 
Can we filter them out, and if so, how? Our idea is to validate them using our trace semantics; see Algorithm~\ref{alg:validate}. 
In a nutshell, given an s-rewrite $\statement'$ corresponding to $\statement_i, \mydots, \statement_j$, the algorithm checks whether it is a true rewrite or not; if so, it returns a slice of statements $\statement_i, \mydots, \statement_r$ in $\program$ that can be rewritten to $\statement'$. 
We require $r \hspace{-2pt}$ > $\hspace{-2pt} j$, because we want $\statement'$ to rewrite a slice of statements beyond $\statement_j$ (i.e., the first iteration). 
Towards this goal, \textsc{Validate} first executes $\statement'$ against the concatenation of DOM traces from $i$ to $l$, yielding an action trace $\actiontrace'$ (line 3). 
Then, line 4 checks if $\statement'$ is a true rewrite; if so, it obtains the rewrite $\program'$ (line 5) and the matching traces (lines 6-7), which are then added to $W$ (line 8). 
Note that invariants $\invariant_1, \invariant_2$ hold for this rewrite $(\program', \vec{\actiontrace}', \vec{\statetrace}')$, as $\actiontrace'$ is obtained by executing $\statement'$ using our trace semantics and is also checked at line 4.

\begin{figure}[!t]
\vspace{-10pt}
\begin{algorithm}[H] 
\footnotesize
\caption{\textsc{Validate} procedure.}
\label{alg:validate}
\begin{algorithmic}[1]

\myprocedureindent{Validate}{$\srewrites, \program, \vecactiontrace, \vecstatetrace$}{-15pt}

\Statex\Input{a set $\srewrites$ of s-rewrites of the form $(\statement', \statement_i, \statement_j)$.} 
\Statex\Input{a program $\program = \statement_1; \mydots; \statement_l$ that each s-rewrite in $\srewrites$ may apply to.} 
\Statex\Input{$\vecactiontrace = [ \actiontrace_1, \mydots, \actiontrace_l ]$, $\vecstatetrace = [ \statetrace_1, \mydots, \statetrace_l ]$.}

\Statex\Output{a set $\rewrites$ of true rewrites of the form $(\program', \vecactiontrace', \vecstatetrace' )$ \emph{s.t.} $\invariant_1,\invariant_2$ hold.}

\State $\rewrites \assign \emptyset$;

\ForAll{$(\statement', \statement_i, \statement_j) \in \srewrites$}

\State $\actiontrace' \assign \textsc{Execute}(  \statement', {\statetrace_i \traceconcat \hspace{-2pt} \cdot \hspace{-2pt} \cdot \hspace{1pt} \traceconcat \statetrace_l}, \inputdata )$; 


\If{$\exists r \in [j+1, l]: \actiontrace' = \actiontrace_i \traceconcat \hspace{-2pt} \cdot \hspace{-2pt} \cdot \hspace{1pt} \traceconcat \actiontrace_r \ \emph{given} \ \statetrace_i \traceconcat \hspace{-2pt} \cdot \hspace{-2pt} \cdot \hspace{1pt} \traceconcat \statetrace_r$}
\State $\program' \assign \block_1; \mydots; \block_{i-1}; \block'; \block_{r+1}; \mydots; \block_{l}$; 
\State $\vec{\actiontrace}' \assign [ \actiontrace_1, \mydots, \actiontrace_{i-1}, \actiontrace', \actiontrace_{r+1}, \mydots, \actiontrace_l ]$; 
\State $\vec{\statetrace}' \assign [ \statetrace_1, \mydots, \statetrace_{i-1}, \statetrace', \statetrace_{r+1}, \mydots, \statetrace_l ]$; 
\State $\rewrites \assign \rewrites \setor \{  ( \program', \vec{\actiontrace'}, \vec{\statetrace'} ) \}$;
\EndIf

\EndFor

\State\Return $\rewrites$;

\end{algorithmic}
\end{algorithm}
\vspace{-22pt}
\end{figure}

\begin{example}
Consider the s-rewrite $(\statement', \action_1, \action_2)$ returned by \textsc{Speculate} in Example~\ref{ex:parametrize}. By construction, the first iteration of $\statement'$ produces $[\action_1, \action_2]$. 
To validate this s-rewrite, we evaluate $\statement'$ against $[\mystate_1, \mydots, \mystate_{60}]$ using our trace semantics, which gives an action trace $\actiontrace' = [ \action_1, \mydots, \action_{40} ]$. This is indeed a true rewrite; thus, \textsc{Validate} returns $\statement'$ together with its matching action trace $[\action_1, \mydots, \action_{40} ]$, indicating $\statement'$ rewrites actions $\action_1, \mydots, \action_{40}$. 
\label{ex:validate}
\end{example}

\subsection{Incremental Synthesis}

Recall from Figure~\ref{fig:workflow} that our synthesizer is used in an iterative fashion: it predicts the next action given the current trace with $m$ actions, where $m$ increases as the task progresses. Therefore, we invoke our synthesis algorithm \emph{incrementally}. 
This is done by simply sharing the worklist in Algorithm~\ref{alg:top} across synthesis runs. 
Suppose we want to synthesize from a trace with $m$ actions, given the worklist $\worklist$ from the previous run. Instead of starting from scratch (line 2, Algorithm~\ref{alg:top}), we resume from $\worklist \cup \worklist'$, where $\worklist'$ contains those programs removed from $\worklist$ (line 4) in the previous run. 
This essentially makes the entire rewrite process across runs not destructive.

\subsection{Soundness and Completeness}

\begin{theorem}
\textnormal{
\hspace{-8pt}
Given action trace $\actiontrace$, DOM trace $\statetrace$ and input data $\inputdata$, if there exists a web RPA program that generalizes $\actiontrace$ and in which every loop has at least two iterations exhibited in $\actiontrace$, then our synthesis algorithm would return a program that generalizes $\actiontrace$ given $\statetrace$ and $\inputdata$. 
}
\label{theorem:thm}
\end{theorem}

%% file: interface.tex
\section{Human-in-the-Loop Interaction Model}
\label{sec:interface}

In this section, we describe our system interaction design rationale and user interface. Our overall design goal is to reduce the gulfs of execution and evaluation for novice users~\cite{norman2013design}. That is, through our interface, we aim to help users better understand what is going on in the system (\ie, evaluation) as well as help them execute intended actions (\ie, execution). 

\noindent
To achieve this goal, we designed a user interaction model that combines PBD and user interaction in a human-in-the-loop process. We highlight some key features below.

\textbf{\em Demo-auth-auto workflow.}
As illustrated in Section~\ref{sec:overview}, there are three phases when using our tool: (a) a \emph{demonstration} phase where the user manually performs a few actions, (b) an \emph{authorization} phase where the user accepts or rejects predictions, and (c) an \emph{automation} phase where our tool automatically executes the program. Our system could transition from one phase to another (automatically or manually).

\textbf{\em Data entry via drag-and-drop.}
Instead of manually typing strings from the input data source, our interface supports drag-and-drop. This design not only simplifies the data entry process but also makes synthesis easier.

\textbf{\em Action highlighting.}
Each action performed on the page is highlighted. In addition, during the demonstration phase, our system also highlights DOM nodes that are hovered over. These designs help users better interact with our system.

\textbf{\em Prediction authorization.}
During the authorization phase, each predicted action requires user approval before it is executed. Our user interface visualizes predictions in an easy-to-examine manner, which helps reduce the gulf of evaluation.

\textbf{\em Navigating across multiple predictions.}
In case there are multiple predictions, our interface will show a navigation arrow which allows users to inspect each of them and accept the desired one, which helps resolve ambiguity.

%% file: eval.tex
\section{Evaluation}
\label{sec:eval}

\begin{figure*}
\centering
\includegraphics[width=.99\textwidth]{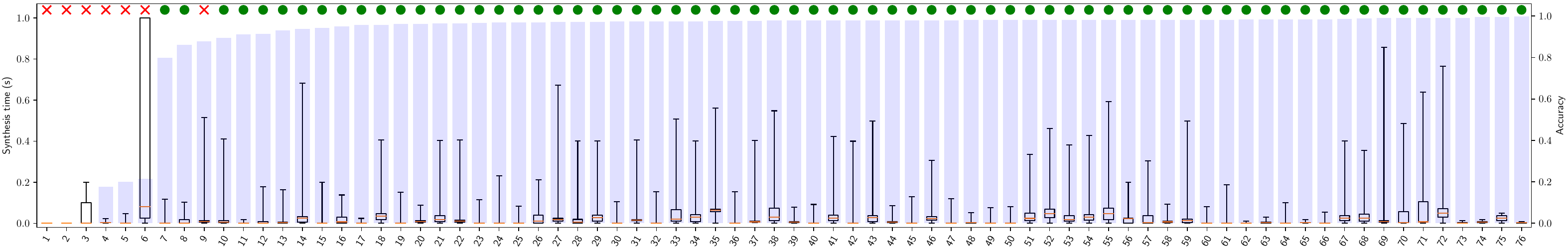}
\vspace{-8pt}
\caption{Main results for Q1: (a) accuracy (bar chart), (b) synthesis time (box plot), and (c) whether the final synthesized programs are intended ({\color{green}{$\bullet$}}/{\bf \color{red}{$\times$}} marks near the top). In particular, benchmarks are sorted in ascending order based on accuracy. For each benchmark, we report quartile statistics of the synthesis times across all tests for which we can produce a prediction within the timeout (1 sec). If the final synthesized program is intended, we mark {\color{green}{$\bullet$}} at the top for that benchmark; otherwise, {\bf \color{red}{$\times$}}.}
\label{fig:time}
\vspace{-10pt}
\end{figure*}

In this section, we describe a series of experiments that are designed to answer the following research questions: 
\begin{itemize}[leftmargin=*]
\item 
\textbf{Q1:} 
Can \toolname's synthesis engine effectively synthesize web RPA programs from demonstrations? 
\item 
\textbf{Q2:}
How important are the ideas proposed in Section~\ref{sec:algorithm}? 
\item 
\textbf{Q3:}
How well does \toolname work \emph{end-to-end} (including both front-end and back-end) in practice? 
\item
\textbf{Q4:}
How does \toolname's performance compare against existing rewrite-based synthesis approaches? 
\end{itemize}

\textbf{\em Implementation.} 
\toolname has been implemented with the proposed ideas as well as several additional optimizations. 
More details can be found in the extended version~\cite{extended}.

\emph{\textbf{Benchmarks.}}
We also constructed a suite of benchmarks for web RPA.
In particular, we first scraped all posts under the \emph{``Data Extraction and Web Screen Scraping''} topic from the iMacros forum. Then, we retained every post that corresponds to a web RPA task with a working URL (e.g., we filter out posts regarding ``how to use iMacros'').

\textbf{\em Ground-truth programs.}
For each benchmark, we have also manually written a program that can automate the corresponding task using the Selenium WebDriver framework. These programs are treated as the ``ground-truth'' programs. 

\textbf{\em Statistics of benchmarks.}
In total, we collected 76 benchmarks with their corresponding ground-truth programs. In particular, all of these benchmarks involve data extraction, 29 of them involve data entry, 60 require navigation across webpages, and 33 involve pagination. Some benchmarks may involve multiple types of actions: for instance, 28 of them involve data entry, data extraction, and webpage navigation. 
The ground-truth programs consist of 36.3 lines of code on average (max being 142). In general, it took us 30 minutes to a few hours to implement a working Selenium program.

\subsection{Q1: Evaluating \toolname's Synthesis Engine}
\label{sec:q1}

Recall that our synthesizer takes as input (1) a demonstrated action trace $\actiontrace = [ \action_1, \mydots, \action_m]$, (2) a DOM trace $\statetrace = [ \mystate_1, \mydots, \mystate_{m+1} ]$, and (3) an optional data source $\inputdata$. 
It returns a program $\program$ that generalizes $\actiontrace$ given $\statetrace$ and $\inputdata$. That is, $\program$ not only reproduces $\actiontrace$ but also predicts a next action $\action_{m+1}$. Thus, our synthesis goal is to generate $\action_{m+1}$ efficiently and accurately.

\textbf{\em Setup.}
To evaluate our synthesis efficiency and accuracy, we designed the following experiment. First, for each benchmark, we instrumented its ground-truth program $\program_{\emph{gt}}$ such that $\program_{\emph{gt}}$ would record every action it executes as well as all intermediate DOMs. Hence, we can obtain the \emph{entire} action trace $\actiontrace_{\emph{gt}} = [ \action_1, \mydots, \action_n ]$ and DOM trace $\statetrace_{\emph{gt}} = [\mystate_1, \mydots, \mystate_n]$\footnote{
We terminate $\program_{\emph{gt}}$ after 500 actions in case it unnecessarily takes long to finish. That is, we may use a prefix of $\program_{\emph{gt}}$'s entire trace in this experiment.}. 
Here, $\action_1$ is the first action performed on $\mystate_1$ and $\action_n$ is the last action on $\mystate_n$. We also ensure that the recorded actions are in the same trace language defined in Section~\ref{sec:dsl}. We convert the recorded selectors used in $\program_{\emph{gt}}$ to \emph{absolute} XPath expressions. The reason is because \toolname's front-end records actions using absolute XPath during user interactions and we aim to simulate that in this experiment. Note that this actually makes synthesis more challenging since we necessarily need to consider alternative selectors in order to synthesize $\program_{\emph{gt}}$. For those benchmarks involving programmatic data entry, we manually constructed a data source $\inputdata$ with $100$ entries.\footnote{Fun fact: we leveraged \toolname when collecting these data sources.}

Given $\actiontrace_{\emph{gt}}$ and $\statetrace_{\emph{gt}}$, we generate $n-1$ tests for the synthesis engine. That is, for the $k$th test, we are given $\actiontrace_k = [\action_1, \mydots, \action_k]$ with the first $k$ actions in $\actiontrace_{\emph{gt}}$ and $\statetrace_{\emph{k+1}} = [\mystate_1, \mydots, \mystate_{k+1}]$ with the first $k+1$ DOMs in $\statetrace_{\emph{gt}}$, and our goal is to synthesize a program that predicts $\action_{k+1}$.
In this setting, we define 
accuracy as the percentage of tests for which we can generate a \emph{correct} prediction that is equivalent to the ground-truth action. 
For efficiency, we calculate the quartile statistics of the synthesis times across all tests that we can generate predictions for. 
In this experiment, we use 1 second as the timeout per test.

\textbf{\em Main results.}
Overall, as shown in Figure~\ref{fig:time}, our synthesis engine solved most benchmarks with both high accuracy and efficiency. In particular, for 68\% of the benchmarks, it achieves at least 95\% accuracy within 0.5 seconds per prediction. 
Furthermore, it generates desired programs for 91\% of the benchmarks. 
We note that it does not need the entire trace (with 500 actions) to generate those desired programs; rather, it typically generalizes with a few dozens of actions (and at most a couple hundreds). Also note that only a very small number of these actions (typically around 10) are \emph{manually} demonstrated. 
Therefore, we believe our synthesis engine can be used in practice to interactively automate web RPA tasks. On average, the final synthesized programs have 6 statements and the largest program has 18. \toolname can also synthesize programs with complex nesting structures: 
32 of them involve doubly-nested loops and 6 involve at least three levels of nesting.  
Thus, we believe our synthesis engine has the potential to scale to complex web RPA tasks.

In what follows, we discuss some interesting findings.

\textbf{\em Ambiguity.} 
The synthesis engine generated multiple programs for 59 of our benchmarks. For 21 of them, it generated multiple predictions. The maximum numbers of synthesized programs and predictions are 101 and 6 respectively. 
This shows that web RPA is a domain with a fair amount of ambiguity, where there could exist multiple semantically different programs satisfying the same specification.

\textbf{\em Pagination beyond ``next page''.}
Some websites use other mechanisms for pagination. For instance, \emph{b9} involves a job search site\footnote{\url{https://www.timesjobs.com/}} 
which performs pagination using page numbers and a ``next 10 pages'' button. We do not support such pagination mechanisms yet. 
The reason \emph{b9} has an 88\% accuracy is because it synthesized a program with a sequence of 
selector loops, which solved the tests but is not intended.

\textbf{\em Complex selectors.}
Some benchmarks need selectors with multiple attributes in order to be automated. For example, \emph{b6} involves scraping players information for matches that have \emph{either ``match'' or ``match highlight'' class}.
Our DSL currently does not support such ``disjunctive logics'' for selectors. Some other benchmarks (such as \emph{b1-3}) also have similar issues.

\textbf{\em Others.}
The reason that \emph{b7} has a relatively low accuracy (80\%), albeit an intended program was synthesized, is because its trace is relatively short (with 51 actions in total) and the intended program was synthesized after the first 10 actions. 
This is also the case for some benchmarks, such as \emph{b8}, \emph{b10-12}.

\subsection{Q2: Ablation Studies of the Synthesis Engine}
\label{sec:ablation}

\textbf{\em Setup.}
We performed ablation studies to quantify the impact of our ideas. In particular, we consider the following variants. 
\vspace{-14pt}
\begin{itemize}[leftmargin=*]
\item 
\emph{No selector}: 
We modified the \emph{AlternativeSelectors} function from Figures~\ref{fig:anti-unify} and~\ref{fig:parametrize} to always return the input selector. Effectively, this variant can only use full XPath expressions from the trace without considering alternative selectors. 
\item 
\emph{No incremental}: This variant does not reuse rewrites from prior synthesis runs. It always starts from scratch if the program $\program_k$ generated for the $k$th test fails to predict $\action_{k+1}$. 
\end{itemize}
\vspace{-2pt}

We conducted the same experiment described in Section~\ref{sec:q1} using these variants. 
Note that we do not include an ablation for the idea of speculative rewriting. The reason is because it is not easy to ``disable'' speculation without fundamentally changing our algorithm. Instead, please see Section~\ref{sec:comparison} for a comparison against a baseline implemented using \texttt{egg}. 

\textbf{\em Main results.}
As shown in Table~\ref{table:q2}, our main take-away is that it is important to perform selector search and incremental synthesis in order to synthesize programs both accurately and efficiently. 
For instance, without considering alternative selectors, it only solves 38 benchmarks and the average accuracy drops to 57\%. 
In terms of synthesis time, all techniques are fairly efficient (for tests that terminate within 1 second). 

\subsection{Q3: Evaluating \toolname End-to-End}

\textbf{\em User study setup.}
To evaluate \toolname end-to-end and access whether it helps users complete web RPA tasks, we conducted a user study involving 8 participants (avg. age 21) from the lead author's institution. All participants are undergraduate students with an average of 4 years of programming experience. Each participant was asked to complete 5 tasks sampled from our benchmarks, divided in three phases. 
\begin{itemize}[leftmargin=*]
\item 
\emph{Phase 1} has one single-page scraping task. 
\item 
\emph{Phase 2} includes two scraping tasks that involve webpage navigation and pagination. 
\item 
\emph{Phase 3} has two tasks that involve data entry. In particular, given a list of keywords, the user needs to perform search on the website using each keyword and then scrape certain information from the search result of each keyword. 
\end{itemize}
For each phase, participants started by watching a tutorial and replicating a demo task. Then, they worked on the main tasks. Each participant had one hour in total for all 5 tasks. 

\textbf{\em User study results.}
All participants were able to successfully automate all tasks using \toolname. In particular, participants demonstrated 6-10 actions before \toolname could automate the rest of the task.
For each phase, the average time it took them to provide demonstrations is (in seconds): 16.88 (SD=3.80, phase 1), 19.44 (SD=11.48, phase 2), and 64.44 (SD=22.58, phase 3). 
Furthermore, all participants were able to use \toolname to resolve ambiguity interactively. Finally, according to a follow-up survey, all eight participants gave positive feedback on the usability of our tool: for instance, they thought \toolname was ``quite decent'' (\emph{P4}), experience was ``smooth'' (\emph{P8}), and it ``can apply to many scenarios'' (\emph{P2}).

\textbf{\em More comprehensive end-to-end testing results.}
To gain a more comprehensive understanding on how \toolname works end-to-end, we tested it on \emph{all} of our 76 benchmarks. While this experiment is inevitably biased because the testers are developers of the tool, we believe it complements the user study and hence is still valuable. 
A benchmark is counted as ``solved'' if we can use \toolname to synthesize the intended program which can automate the benchmark.\footnote{For long-running programs (>10 minutes), we ran them for 3 iterations.} 
Overall, we solved 76\% of the benchmarks by interactively demonstrating around 10 actions. We also found it necessary to resolve ambiguity when solving these benchmarks. 
Among the remaining 18 benchmarks, 7 failed due to the issues from \toolname's back-end (see Section~\ref{sec:q1}), and 11 were due to limitations of our front-end. For instance, our current front-end does not fully support replaying certain actions in a few situations, which accounts for 7 cases.

\textbf{\em Discussion.}
Comparing the end-to-end testing conducted by ourselves (i.e., \toolname developers) and the user study with novice users, a notable difference in use patterns is that novice users make mistakes (e.g., mis-clicks on webpages and mis-use of the UI). In this case, we assisted the participants to restart \toolname and perform the task again.

\begin{table}[!t]
\caption{\small Main results for ablation studies in Q2.}
\vspace{-10pt}
\scriptsize
\centering
\begin{centermath}
\begin{array}{l}
\setlength{\tabcolsep}{3pt}
\begin{tabular}{l  | R{42pt} R{30pt} R{30pt} R{40pt} }
\hline 
& \# Benchmarks & Accuracy & Accuracy & Time per test   \\[-2pt]
Variants \hspace{7pt} & solved &  (median) & (average) &  (average) \\[1pt] \hline \\[-6pt]
\emph{Full-fledged}  \hspace{7pt} & 69 & 98\% & 90\%  & 23ms \\ 
\emph{No selector} & 38 & 88\% & 57\% & 54ms  \\ 
\emph{No incremental} & 45 & 96\% & 72\% & 32ms \\ \hline 
\end{tabular}
\end{array}
\end{centermath}
\label{table:q2}
\vspace{-10pt}
\end{table}

\subsection{Q4: Comparison with Existing Rewrite-Based Synthesis Techniques}
\label{sec:comparison}

The goal of this final experiment is to understand how our speculative rewriting idea compares with existing rule-based rewrite approaches that perform synthesis in a correct-by-construction manner (without speculation). 
Specifically, we implemented a baseline synthesizer using the \texttt{egg} library~\cite{willsey2021egg}, a state-of-the-art framework that was used to build many high-performance rewrite-based synthesizers~\cite{vanhattumvectorization,yang2021equality,panchekha2015automatically,nandi2020synthesizing}.

\textbf{\em Our \egg-based implementation.} 
Our baseline consists of two key rules: one that splits a trace into slices and another that ``rerolls'' a slice into a loop. 
We illustrate these rules using Example~\ref{ex:challenge}, by showing one specific sequence of rules that rewrites $\actiontrace$ to $\program$. 
Conceptually, we first apply a \emph{Split} rule that splits $\actiontrace$ to three slices: 
\[
[ \action_1, \mydots, \action_{59} ] 
\rightarrow
\emph{Unsplit}(
[ [ \action_1, \mydots, \action_{40} ], [ \action_{41} ], [ \action_{42}, \mydots, \action_{59} ] ]
)
\] 
Then, we use a \emph{Reroll} rule that yields two rewrites:
\[
\begin{array}{c}
[ \action_1, \mydots, \action_{40} ] 
 \ \rightarrow
\emph{InnerLoop}   
\end{array}
\begin{array}{c}
[ \action_{42}, \mydots, \action_{59} ] 
\rightarrow
\emph{InnerLoop}   
\end{array}
\]
In other words, these two slices are rewritten to two instances of the inner loops of $\program$. 
Note that now the e-graph contains $\emph{Unsplit}([ [\emph{InnerLoop}], [\action_{41}], [\emph{InnerLoop}] ])$. 
The third rule we apply is \emph{Unsplit} which does the following rewrite: 
\[
\hspace{-50pt}
\begin{array}{c}
\emph{Unsplit}([ [\emph{InnerLoop}], [\action_{41}], [\emph{InnerLoop}] ]) 
\rightarrow \\ 
\hspace{170pt} [ \emph{InnerLoop}, \action_{41}, \emph{InnerLoop} ]
\end{array}
\]
As one can imagine, we can apply the aforementioned rules again to finally generate $\program$. While our current baseline only supports selector loops without alternative selectors, it leverages e-class analysis and is fairly optimized. Thus, we believe it is still a good baseline to test the performance of a purely rule-based, correct-by-construct synthesis approach.

\textbf{\em Results and discussion.} 
We evaluated this baseline on \emph{all} 9 benchmarks whose ground-truth programs involve only selector loops and no alternative selectors. In particular, we ran it on action traces of increasing length (from length 1). Table~\ref{table:q3egg} presents the synthesis time for the shortest trace, for which it gives an intended program, across \emph{all} benchmarks. 
Our main take-away is that the baseline does not scale well. In particular, it solved 7 tasks whose ground-truth programs all have one single loop. \emph{b12}, which requires synthesizing a doubly-nested loop, took 200s. The most complex problem is \emph{b56}, which involves a three-level loop, and it did not terminate in 5 minutes. On the other hand, \toolname solved all 9 benchmarks with at most 1 second.

\begin{table}[!t]
\caption{\small Main results for \texttt{egg}-based implementation in Q4. X/Y gives synthesis time X (milliseconds) for (shortest) trace length Y.}
\vspace{-10pt}
\em 
\scriptsize
\centering
\begin{centermath}
\begin{array}{l}
\hspace{-4pt}
\setlength{\tabcolsep}{3pt}
\begin{tabular}{l  | rrrrrrrrr }
\hline \\[-7pt] 
& {b12} & {b15}  &  {b20} & {b48} & {b56} & {b73} & {b74} & {b75} & {b76}  \\ \hline \\[-6pt]
Baseline using \egg & \emph{2$\times10^5$/34}  &  \emph{12/6} & \emph{15/12} & \emph{6/8} & \emph{--/--} & \emph{2/2} & \emph{2/2} & \emph{3/2} & \emph{2/2}  \\[1pt]
\toolname & \emph{186/34} & \emph{11/6} & \emph{22/12} & \emph{12/8} & \emph{950/204} & \emph{7/2} & \emph{6/2} & \emph{7/2} & \emph{6/2} \\ \hline 
\end{tabular}
\end{array}
\end{centermath}
\label{table:q3egg}
\vspace{-10pt}
\end{table}

%% file: related.tex
\section{Related Work}
\label{sec:related}

In this section, we briefly discuss some closely related work.

\emph{\textbf{RPA.}}
As a relatively new topic, there is little work on RPA until very recently~\cite{leno2021discovering,leno2020robidium,zhang2020process,wewerka2020robotic,agostinelli2020towards}. Existing work mainly focuses on formalizing key concepts and the routine discovery problem. 
In contrast, our work targets a different problem of how to help non-experts create automation programs, which is also critical for fostering RPA adoption.

\emph{\textbf{Web automation.}}
Similar to web automation, \toolname emulates user interactions with a web browser and hence can be viewed as a form of web automation. Our work differs from prior web automation work~\cite{chasins2018rousillon,barman2016ringer,chasins2015browser,leshed2008coscripter,lin2009end,fischer2021diy,selenium,imacros,cypress-studio,little2007koala,chasins2019democratizing} in several ways. 
One notable difference is that \toolname is based on interactive PBD whereas prior techniques are either program-centric or programmer-centric.

\emph{\textbf{Interactive program synthesis.}}
Different from prior approaches~\cite{kandel2011wrangler,barowy2015flashrelate,le2014flashextract,gulwani2011automating,ferdowsifard2021loopy,wang2019visualization}, which are mostly interactive \emph{programming-by-example}, \toolname is based on interactive \emph{programming-by-demonstration} which is a natural approach for web RPA. While action traces can be viewed as a form of ``examples'', it introduces a new challenge in how to define the correctness of a program against a trace. We use a form of trace semantics for our language, which lays the formal foundation for web RPA program synthesis. 

\textbf{\em Programming-by-demonstration (PBD).}
Existing PBD techniques can be roughly categorized into two groups: those that reason about \emph{user actions} (e.g., \helena~\cite{chasins2018rousillon} and TELS~\cite{mo1990learning}) and those that examine \emph{application states} (e.g., Tinker~\cite{lieberman1993tinker} and SMARTedit~\cite{lau2003programming}). 
While almost all of them rely on heuristic rules to generalize programs from demonstrations~\cite{lau2001programming}, a notable exception is SMARTedit, which proposes a principled approach based on version space algebras that could generate programs from a short trace of \emph{states.} 
Similarly, our work contributes a principled approach but for \emph{action-based trace generalization}. In particular, we propose a rewrite-based algorithm for synthesizing programs from a trace of actions. 

\textbf{\em Term rewriting.}
Term rewriting~\cite{dershowitz1990rewrite} has been used widely in many important domains~\cite{boyle1997tampr,visser1998building,joshi2002denali,tate2009equality,panchekha2015automatically,nandi2020synthesizing,smith2019program,claver2021regis,willsey2021egg,premtoon2020semantic}. 
Our work explores a new application for synthesizing web RPA programs from traces. 
Different from existing rewrite-based synthesis techniques~\cite{willsey2021egg,nandi2020synthesizing} which are mostly purely rule-based and correct-by-construction, we leverage the idea of guess-and-check in the overall rewrite process. This new speculate-and-validate methodology enables and accelerates web RPA program synthesis.

\textbf{\em Program synthesis with loop structures.}
\toolname is related to a line of work that aims to synthesize programs with \emph{explicit} loop structures~\cite{shi2019frangel,chasins2018rousillon,pailoor2021synthesizing,ferdowsifard2021loopy}. The key distinction is that we use demonstrations as specifications, whereas prior approaches are mostly programming-by-example.

\textbf{\em Human-in-the-loop.}
\toolname adopts a human-in-the-loop interaction model which has shown to be an effective way to incorporate human inputs when training AI systems in the HCI community~\cite{nushi2017human,chen2020bashon}. 
This model has been used in the context of program synthesis, mostly interactive PBE~\cite{newcomb2019using,zhang2020interactive,wang2021falx,naik2021sporq,ferdowsifard2020small,santolucito2019live}. In contrast, our work incorporates human inputs in PBD and proposes a new human-in-the-loop model.

%% file: ack.tex
\vspace{-2pt}
\begin{acks}                            
We thank our shepherd, Calvin Smith, the PLDI anonymous reviewers, Kostas Ferles, Cyrus Omar, Shankara Pailoor, Chenglong Wang, and Yuepeng Wang  for their feedback. 
We also thank Yiliang Liang and Minhao Li for their help with our \egg baseline implementation.
This work was supported by the National Science Foundation under Grant No. 2123654. 
\end{acks}